\documentclass[12pt]{article} % For LaTeX2e
\newcommand{\includeappendix}{} % Comment out to exclude appendix
\usepackage{rldmsubmit, palatino}
\usepackage{graphicx, wrapfig}
\usepackage[backend=biber, style=apa, natbib=true]{biblatex}
\usepackage{amsmath, amsfonts}
\usepackage[newfloat,frozencache,cachedir=_minted-main]{minted}
\ifdefined\includeappendix\else
  \usepackage{xr}
\fi

\usepackage{caption}
\usepackage{tikz}
\usetikzlibrary{positioning}
\usepackage{tikz-3dplot}
\usepackage{tikzpeople}
\usepackage{adjustbox}
\usepackage{subcaption}
\usepackage{booktabs} % For formal tables
\usepackage{array}    % For table wrapping
\usepackage{longtable}% For tables that span multiple pages
\usepackage{xcolor}
\usepackage{tcolorbox}
\tcbuselibrary{minted, skins, listingsutf8, breakable}
\usepackage{setspace}
\usepackage{xspace}
\usepackage{fontawesome5}

\newcommand{\cogrid}{{\textsc{CoGrid}}\xspace}
\newcommand{\mug}{{\textsc{MUG}}\xspace}
\newcommand{\pyinline}[1]{\mintinline{python}{#1}}

\newcommand{\red}[1]{#1}

\AtBeginEnvironment{minted}{\singlespacing}

\SetupFloatingEnvironment{listing}{%
  name={Listing},
  fileext=lol}

\newtcblisting{mintedbox}[2][]{
  listing engine=minted,
  colback=gray!10, % Background color of the box
  colframe=black, % Color of the frame
  listing only,
  minted language=#2,
  minted options={linenos=true, breaklines=true, fontsize=\footnotesize, numbersep=3mm, baselinestretch=1},
  left=5mm, % Left margin within the box
  enhanced,
  breakable,
  toprule at break=0pt,
  bottomrule at break=0pt,
  #1 % For additional options
}

\newsavebox{\IGConfig}
\begin{lrbox}{\IGConfig}
\begin{adjustbox}{minipage=0.40\textwidth}
\begin{minipage}{0.5\textwidth} % Adjust the width as necessary
\begin{minted}{python}
config = interactive_gym.Config()
    .policies(...)
    .environment(...)
    .rendering(...)
    .gameplay(...)
    .hosting(...)
    .user_experience(...)

if __name__ == "__main__":
    app.run(config)
\end{minted}
\end{minipage}
\end{adjustbox}
\end{lrbox}

\newsavebox{\OvercookedEnv}
\begin{lrbox}{\OvercookedEnv}
\begin{adjustbox}{minipage=0.43\textwidth}
\begin{minipage}{0.5\textwidth} % Adjust the width as necessary
\begin{minted}{python}
class Overcooked(CoGridEnv):
    def __init___(*args, **kwargs):
        [...]

class Onion(GridObj):
    object_id = "onion"
    color = "yellow"

    def __init__(*args, **kwargs):
        [...]

[...]
\end{minted}
\end{minipage}
\end{adjustbox}
\end{lrbox}

\newsavebox{\OvercookedConfig}
\begin{lrbox}{\OvercookedConfig}
\begin{adjustbox}{minipage=0.5\textwidth}
\begin{minipage}{0.5\textwidth} % Adjust the width as necessary
\begin{minted}{python}
config = interactive_gym.Config()
    .policies(
        policy_mapping={
        "agent_0": Human,
        "agent_1": "model.onnx",
        }
        [...]
    )
    .environment(
        env_creator=lambda: Overcooked(),
    )
    .rendering(
        env_to_render_fn=render_fn,
        [...]
    )
    .gameplay([...])
    .hosting([...])
    .user_experience([...])

if __name__ == "__main__":
    app.run(config)
\end{minted}
\end{minipage}
\end{adjustbox}
\end{lrbox}

\newsavebox{\OvercookedLandingPage}
\begin{lrbox}{\OvercookedLandingPage}
\begin{minipage}{0.5\textwidth} % Adjust the width as necessary
    \includegraphics[width=\textwidth]{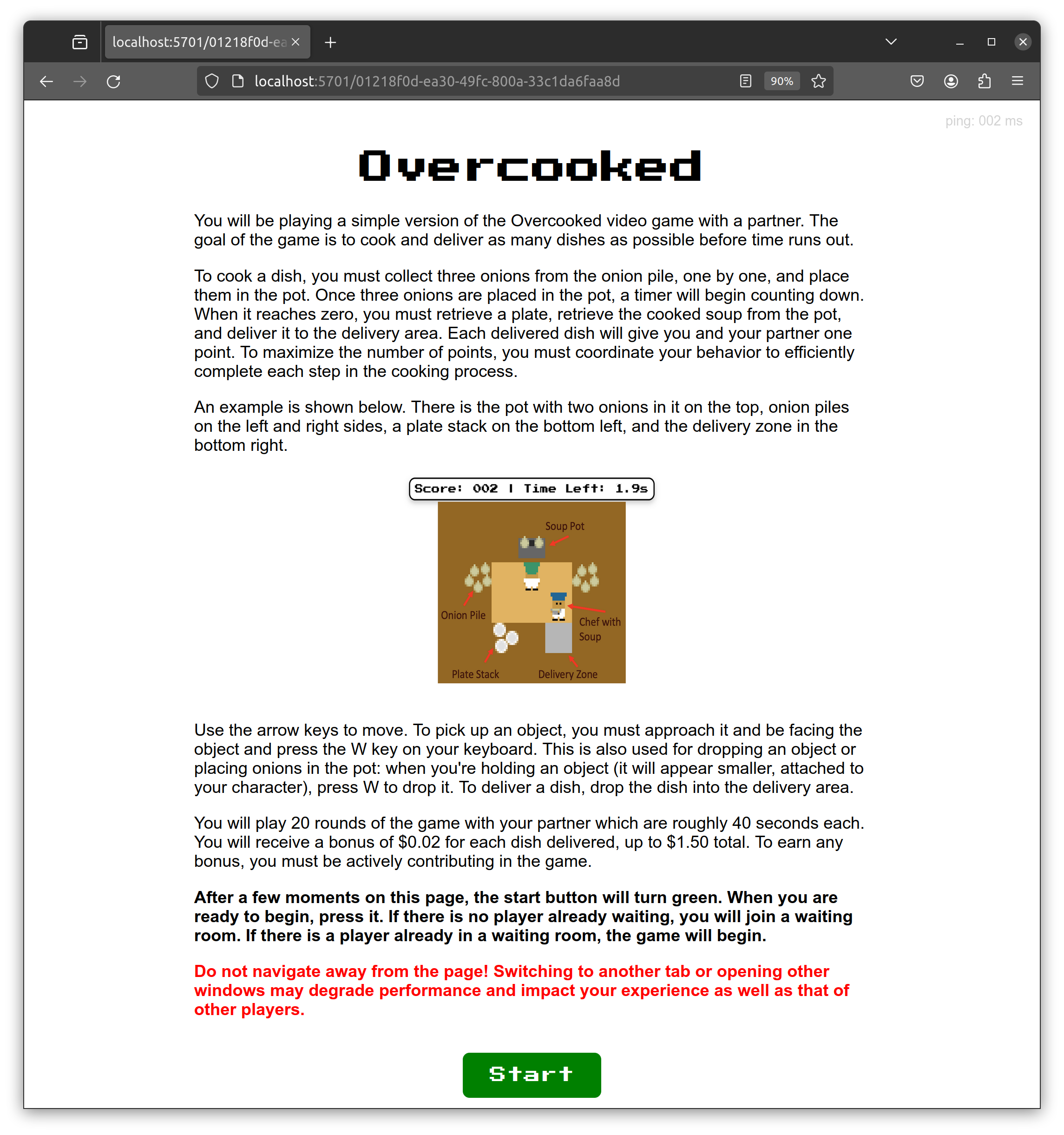}
\end{minipage}
\end{lrbox}

\newsavebox\Server
\sbox\Server{\tdplotsetmaincoords{70}{20}
\begin{tikzpicture}[tdplot_main_coords]
 \begin{scope}[canvas is xz plane at y=3]
  \path (1,0)  coordinate(aux);
  \path (1pt,0) coordinate (BTL) (1cm-1pt,0) coordinate (BTR)
  (1cm,-2.5cm+1pt) coordinate (BBR) (1cm,-1pt) coordinate (BTR');
 \end{scope}
 \begin{scope}[canvas is xz plane at y=0]
  \draw[rounded corners={2*sqrt(2)*1pt},fill=gray!10] (0,-2.5) rectangle (1,0);
  \draw[rounded corners={2*sqrt(2)*1pt},white] (0.4pt,-2.5cm+0.4pt) 
    rectangle (1cm-0.4pt,0-0.4pt);
  \path (1pt,0) coordinate (FTL) (1cm-1pt,0) coordinate (FTR);
  \path[fill=white,rounded corners={2*sqrt(2)*1pt}]
   ($(aux)+(-1,0)$) -| ++(1,-2.5) -- (1,-2.5) |- (0,0)--cycle;
  \path[left color=gray!10,right color=gray!30,rounded corners=1pt] (BTL) -- (BTR) -- (FTR) -- (FTL)
    -- cycle;
  \path[top color=gray!80,bottom color=gray!30,shading angle=20,
    rounded corners=1pt]
   (1cm,-2.5cm+1pt) -- (BBR) -- (BTR') -- (1cm,-1pt) -- cycle;
  \draw[ultra thin,fill=gray!40] foreach \X in {0.2,0.3,...,0.81} 
  { \foreach \Y in {-0.2,-0.3,...,-2.3} 
   {(\X-0.03,\Y-0.03) rectangle (\X+0.03,\Y+0.03)}};
  \begin{scope} 
   \clip (0,-2.5) rectangle (1,-0.6pt);
   \fill[gray!10] (0.5,0) circle[radius=0.35cm];
   \shade[ball color=black!80] (0.5,0) circle[radius=0.25cm];
  \end{scope}
  \begin{scope}[rounded corners=1mm] 
   \clip (0.42,-0.7) -- (0.42,-0.9) -- (0.22,-1.1)
    -- (0.42,-1.3) -- (0.42,-2.1) -- (0.58,-2.1) 
    -- (0.58,-1.3) -- (0.78,-1.1) -- (0.58,-0.9) -- (0.58,-0.7) -- cycle;
   \fill[gray!80] (0.42,-0.7) -- (0.42,-0.9) -- (0.22,-1.1)
    -- (0.42,-1.3) -- (0.42,-2.1) -- (0.58,-2.1) 
    -- (0.58,-1.3) -- (0.78,-1.1) -- (0.58,-0.9) -- (0.58,-0.7) -- cycle;
   \fill[gray!20] (0.03+0.42,-0.7) -- (0.03+0.42,-0.9) -- (0.03+0.22,-1.1)
    -- (0.03+0.42,-1.3) -- (0.03+0.42,-2.1) -- (0.03+0.58,-2.1) 
    -- (0.03+0.58,-1.3) -- (0.03+0.78,-1.1) -- (0.03+0.58,-0.9) -- (0.03+0.58,-0.7) -- cycle;       
  \end{scope} 
  \shade[ball color=black!80] (0.5,-1.1) circle[radius=0.1cm];
 \end{scope}
\end{tikzpicture}
}

\usepackage{enumitem,amssymb}
\usepackage{algorithm, algpseudocode}
\usepackage{istgame}
\usetikzlibrary{bayesnet, arrows, backgrounds, calc}
\usetikzlibrary{arrows,decorations.pathmorphing,decorations.pathreplacing,backgrounds,positioning,fit,petri,shapes.geometric,shapes.symbols,fit,positioning,shadows}
\newlist{todolist}{itemize}{2}
\setlist[todolist]{label=$\square$}
\usepackage{pifont}
\newcommand{\cmark}{\ding{51}}%
\newcommand{\xmark}{\ding{55}}%

\title{CoGrid \& the Multi-User Gymnasium: A Framework for Multi-Agent Experimentation}

\author{
Chase McDonald \\
Department of Social and Decision Sciences\\  Carnegie Mellon University \\
\And
Cleotilde Gonzalez \\
Department of Social and Decision Sciences\\  Carnegie Mellon University \\
\texttt{coty@cmu.edu} \\ 
}

\begin{document}

\maketitle
\doublespacing

\begin{abstract}
    The increasing integration of artificial intelligence (AI) in everyday life brings with it new challenges and questions for regarding how humans interact with autonomous agents. Multi-agent experiments\red{, where humans and AI act together,} can offer important opportunities to study social decision making, but there is a lack of accessible tooling available to researchers to run such experiments. We introduce two \red{tools} designed to reduce these barriers. The first, \cogrid, \red{is a multi-agent grid-based simulation library with dual NumPy and JAX backends.} The second, \red{Multi-User Gymnasium (\mug)}, translates such simulation environments directly into interactive web-based experiments. \red{\mug supports interactions with arbitrary numbers of humans and AI, utilizing either server-authoritative or peer-to-peer networking with rollback netcode to account for latency.} Together, these tools can enable researchers to deploy studies of human-AI interaction, facilitating inquiry into core questions of psychology, cognition, and decision making \red{and their relationship to human-AI interaction.} Both tools are \red{open source} and available to the broader research community. Documentation and source code is available at \red{\texttt{\{cogrid, multi-user-gymnasium\}.readthedocs.io}}. This paper details the functionality of these tools and presents several case studies to illustrate their utility in human-AI multi-agent experimentation.
\end{abstract}

% \keywords{}

\acknowledgements{This research was supported by the Defense Advanced Research Projects Agency and was accomplished under Grant Number W911NF-20-1-0006 and by the NSF AI Institute for Societal Decision Making (AI-SDM), Award No. 2229881.}

\pagebreak

% \startmain % to start the main 1-4 pages of the submission.
\section{Introduction}

Artificial intelligence (AI) research has made undeniable progress in producing highly capable systems, yet most benchmarks and evaluations emphasize isolated performance rather than interaction with humans. In real-world settings, AI must not only be competent but also interact effectively with humans---its main beneficiaries---and make decisions that align with and complement human decision-making processes \citep{wilder2020learning, russell2019human}.

Much of the literature has focused on agents that outperform humans in games or on established benchmarks (e.g., \citet{silver_reward_2021, perolat2022mastering, meta2022human}), with far less attention to how such agents interact with or impact human collaborators. Even when human interaction is considered, state-of-the-art AI systems are not always incorporated. For example, research on human-AI teaming often relies on surveys, rule-based expert systems, or Wizard-of-Oz paradigms rather than fully autonomous, learning-based agents \citep[e.g.,][]{zhang2021ideal, salikutluk2024evaluation, schelble2022let, duan2024understanding}. Although early work highlights the importance of understanding human perceptions, preferences, and coordination dynamics \citep{carroll_utility_2020, strouse_collaborating_2021, schmutz2024ai}, progress remains constrained by the lack of accessible platforms for conducting controlled human-AI experiments.

Studying human-AI interaction poses unique challenges compared to pure simulation or human-only studies. In the context of reinforcement learning, designing \red{custom} settings for single agents in pure simulation can be cumbersome \citep{bamford_griddly_2022}. Adding human interaction further complicates such inquiries \citep{ouyang2022training}, requiring the implementation of necessary infrastructure, from networking to user interfaces and data collection pipelines that are often non-\red{standard} for complex tasks and rebuilt from scratch for each study. These are significant barriers to entry in empirical human-AI research.

To address these challenges, we present a framework comprised of two complementary tools. The first, \cogrid, \red{is a multi-agent grid-based simulation library originally inspired by} \textit{Minigrid} \citep{chevalier-boisvert_minigrid_2023}. \red{The library is designed to facilitate extensibility and customization through modularized components.} By adopting the \textit{PettingZoo} API \citep{terry2021pettingzoo}, \cogrid allows researchers to build cooperative or competitive environments that fit into standardized simulation infrastructure (e.g., reinforcement learning algorithm libraries). \red{\cogrid is built to follow the PettingZoo API using a NumPy~\citep{harris2020array} backend; however, it is also built with an optional JAX \citep{jax2018github} backend, enabling hardware acceleration for fast parallelized simulations.}

The second tool, \mug, streamlines the deployment of simulation environments into interactive web-based experiments. \mug \red{takes} Python-based environments \red{that follow the Gymnasium or PettingZoo APIs and makes them playable in the browser} without the need for game engines or following the video game development cycle \citep{glazer2015multiplayer}. \red{Environments can run on the server or directly in participants' browsers via Pyodide~\citep[Python compiled to WebAssembly;][]{pyodide_2021}. For multiplayer experiments, it enables peer-to-peer networking with GGPO-style~\citep[Good Game Peace Out;][]{cannon2019ggpo} rollback netcode for latency correction. \mug provides common experiment infrastructure}---including landing pages, waiting rooms, \red{matchmaking,} data collection, AI inference, and surveys---\red{so that researchers do not need to rebuild these components for each study.}

In this paper, we detail the functionality of these tools, which we provide to the research community at \verb|{cogrid, multi-user-gymnasium}.readthedocs.io|. We also present several case studies to demonstrate the use of both tools for designing simulation experiments and conducting both human-AI and human-human interaction experiments. In summary, our contributions are as follows:

\begin{enumerate}
    \item We release \cogrid, a \red{multi-agent grid-based simulation library with dual NumPy/JAX backends and the PettingZoo API, supporting both rapid prototyping and hardware-accelerated training.}
    \item We release \mug, a \red{platform that deploys Gymnasium and PettingZoo environments as browser-based experiments, with support for client-side execution and peer-to-peer multiplayer.}
    \item We demonstrate the use of both \cogrid and \mug, illustrating how they can facilitate new lines of research in human-AI interaction.
\end{enumerate}
\section{Related Work}

In recent years, there has been extensive development of simulation environments and platforms for behavioral research. In this section, we review the relevant prior work that motivates and complements both \cogrid and \mug.

Minigrid has become a standard for reinforcement learning experiments. Despite the simplicity of grid-based environments, significant complexity can arise in task formulation and dynamics \citep{bamford_griddly_2022, chevalier-boisvert_minigrid_2023}, from social dilemmas \citep{agapiou2022melting} to difficult exploration and generalization tasks \citep{hafner2021benchmarking}. The current work on \cogrid is largely motivated by the accessibility and ease of use of Minigrid \citep{chevalier-boisvert_minigrid_2023}: \cogrid aims to provide an equally accessible multi-agent library for developing simulation environments\red{, with the additional requirement of supporting hardware-accelerated training through a dual backend}. The primary goal is to provide a platform that can be easily manipulated, extended, and customized to ask novel questions, rather than serving as a platform for standardized benchmarks.

\mug aims to bridge the gap between environments that are useful for training and evaluating AI in simulation, and tasks that can be used in human behavioral experiments. While there are several libraries that can be used to create environments, including \cogrid, \red{few generalized platforms exist for translating a standard-API Python-based simulation environment into an interactive experiment \citep{aydin2025sharpie}. \mug addresses this by allowing researchers to deploy simulation environments as browser-based experiments, with environments running either on the server or directly in participants' browsers.}

\subsection{Multi-Agent Environments and Experiments}

The standard interface for the agent-environment cycle in reinforcement learning is Gymnasium \citep{towers_gymnasium_2023}. It defines the API for how an agent can interact with the environment; however, it does not standardize or even allow for multi-agent interactions. The vast majority of multi-agent environments augment the Gymnasium API through minor extensions or alterations, maintaining most of the interface \citep[i.e.,][]{liang2018rllib, terry2021pettingzoo, bamford_griddly_2022}. 

Whereas the development of single-agent environments has largely unified around the Gymnasium API, the adoption of a multi-agent API in the reinforcement learning research community has been less uniform, although several platforms have made strides toward a standard approach. In an extension of the Gymnasium API, the PettingZoo interface \citep{terry2021pettingzoo} has seen significant adoption. PettingZoo provides a standardized API for the agent-environment cycle in multi-agent settings, alongside a number of existing benchmark environments, such as the multi-agent Atari suite.

An API for interacting with an environment is only one component of a multi-agent simulation framework. The other critical component is the implementation of the environment itself, including the state representation and transition dynamics. In the single agent setting, a wide range of mature frameworks and platforms support the construction and extension of environments (e.g., \cite{samvelyan2021minihack, sukhbaatar2015mazebase, chevalier-boisvert_minigrid_2023}). In contrast, the multi-agent domain offers substantially fewer such resources. Existing platforms typically define their own environment-creating formats, which vary considerably in complexity, abstraction level, and degree of customizability.

The framework most closely related to the present work is \textit{Griddly} \citep{bamford_griddly_2022}. Griddly is an open-source gridworld game engine that uses its own description language, Griddly Description YAML (GDY), to configure environments and the interactions within them. It is highly extensible and supports a variety of agent interactions---including both single- and multi-agent. In the present work, we make similar concessions as those made by \citet{chevalier-boisvert_minigrid_2023}: Griddly provides increased functionality relative to both Minigrid and \cogrid. However, this increases the barrier to entry for understanding the library and developing new environments. In addition to the GDY description language for configuration, Griddly relies on a C++ core game engine, trading off \red{ease of use} and readability for efficiency. In the approach we take here, we focus on lowering the barrier to entry to allow researchers to develop environments without needing significant outside knowledge. 

\textit{Melting Pot} \citep{agapiou2022melting} provides an additional framework for grid-based multi-agent environments. Its goal is to provide a tool to help in the development and evaluation of agents in multi-agent environments and their ability to learn policies that generalize to novel partners. Melting Pot includes social dilemma environments utilized in previous work \citep{leibo2017multi, hughes2018inequity, jaques2019social}, which were used to study learning dynamics and behavior with reinforcement learning agents in temporally extended variations of classic economic games (e.g., prisoner's dilemma and tragedy of the commons). Melting Pot relies on a combination of the Lua programming language and Python for the development of environments, resulting in an increased barrier to entry and a higher level of complexity for developing custom environments. 

Aside from generalized platforms for building custom environments, there are a number of specialized frameworks and one-off implementations that provide examples of desired functionality. These range from relatively simple abstractions \citep[e.g.,][]{skrynnik_pogema_2022} to complex 3D video games \citep[e.g.,][]{ellis2024smacv2, gym_derk}.\footnote{A number of such examples can be found through the third-party environments linked in the PettingZoo documentation:  \url{https://pettingzoo.farama.org/environments/third_party_envs/}.} A particularly relevant example is the Overcooked-AI environment developed by \citet{carroll_utility_2020}. In their work, they adapted a popular collaborative video game, Overcooked, into a reinforcement learning environment to construct and evaluate agents for human-AI collaboration. Their setting has inspired a significant amount of follow-up work, with reimplementations and extensions of the Overcooked environment (for example, \cite{agapiou2022melting}). The \red{ad hoc} nature of these implementations also demonstrates the difficulty in designing general-use multi-agent environments, a difficulty that \cogrid is designed to alleviate. 

\red{In the same vein as \cogrid, there} are existing efforts to add multi-agent functionality to Minigrid \citep{ndousse_marlgrid_2020, gym_multigrid}. Despite their initial efforts to provide multi-agent environments, these projects remain unmaintained and undocumented, preventing wider adoption, extension, and customization. Our work provides a complete and documented extension of Minigrid with increased customizability through modularization, as described below.

\red{A separate line of work has addressed the computational limitations of standard simulation environments through hardware acceleration. NAVIX \citep{pignatelli2024navix} reimplements Minigrid entirely in JAX, achieving substantial speedups, but retains Minigrid's single-agent design. JaxMARL \citep{flair2024jaxmarl} extends this idea to the multi-agent setting, implementing multi-agent environments and algorithms entirely in JAX for hardware-accelerated training. However, both NAVIX and JaxMARL are JAX-only, which limits their applicability in settings that require web-compatible environments for browser-based execution. \cogrid takes a different approach through its dual-backend architecture: environments are written against a shared array namespace that dispatches to either NumPy or JAX at runtime, enabling the same environment code to be used for both rapid prototyping and high-throughput training. This allows \cogrid to utilize the JAX backend for accelerated training, then deploy the exact same environment to run natively in browsers using the NumPy backend in \mug.} \red{Table~\ref{tab:env_comparison} summarizes the key dimensions along which the aforementioned environment libraries differ.}

\red{
  \begin{table}[htbp]                                                                                                                            
  \centering                                                                                                                                     
  \footnotesize                                                                                                                                  
  \begin{tabular}{l c c c c c}                                                                                                                   
  \toprule                                                                                                                                       
   & Language & Multi-agent & Hardware Acceleration & Standard API & Web Compatible \\                                                           
  \midrule
  Minigrid & Python & \xmark & \xmark & Gymnasium & \xmark \\
  NAVIX & Python & \xmark & \cmark & Gymnasium & \xmark \\
  Griddly & GDY + C++ & \cmark & \xmark & Custom & \cmark \\
  Melting Pot & Lua + Python & \cmark & \xmark & Custom & \xmark \\
  JaxMARL & Python & \cmark & \cmark & Custom & \xmark \\
  \cogrid & Python & \cmark & \cmark & PettingZoo & \cmark \\
  \bottomrule
  \end{tabular}
  \caption{\red{Comparison of RL environment libraries. \emph{Language} indicates the language(s) required to implement new environments.
  \emph{Multi-agent} indicates native support for multi-agent settings. \emph{Hardware Acceleration} indicates support for hardware-accelerated
  (GPU/TPU) training. \emph{Standard API} indicates compatibility with Gymnasium or PettingZoo. \emph{Web Compatible} indicates the ability to
  run in a web browser (e.g., via WebAssembly compilation).}}
  \label{tab:env_comparison}
  \end{table}
}

\subsection{Human-AI Interaction and Interactive Platforms}

In addition to training and testing autonomous agents in a pure simulation environment, there have been several efforts to provide platforms that enable humans to interact with these agents. Each platform has been developed largely to fill a specific need in the research community---from interactive versions of single games \citep{carroll_utility_2020} to frameworks for decision-making research \citep{balietti2017nodegame}---and they highlight the growing amount of research into human-AI interaction. However, existing work has identified that there is a gap for generalized platforms that take complex tasks used in simulation and allow for human interaction \citep{aydin2025sharpie}. We provide an overview of some of the existing approaches and the particular niche that they fill.

A generalized framework for real-time games and experiments with human participants is \textit{nodeGame} \citep{balietti2017nodegame}. It provides researchers with a JavaScript framework to design games for individual human participants, multiple humans, and human-bot populations. It provides a framework that streamlines the creation of experiment flow, participant interaction, data collection, and much more. However, to implement more complex real-time games, as are commonly used in multi-agent reinforcement learning, the onus remains on the developer to integrate the game logic into the nodeGame experiment flow. Indeed, nodeGame is primarily used in settings where participants interact by selecting between alternatives (e.g., Prisoner's Dilemma) or making allocations (e.g., Dictator Game), among other paradigms common in behavioral research.

A particularly relevant development that inspired the setup and laid the groundwork for \mug is the interactive \textit{Overcooked-AI} of \cite{carroll_utility_2020}. In their work, they provide an interactive demonstration of their Overcooked environment, which can be used to play the game with any combination of human and AI agents. Their framework is designed specifically for their implementation of Overcooked-AI, rather than being extensible to different environments. This lack of generality is exactly what motivated the development of \mug: we enhance their approach to provide a mapping from a generic class of simulation environments to interactive tasks\red{, embedded within the full experiment pipeline---including surveys, condition randomization, instructions, and more}. 

\red{A recent and closely related effort is SHARPIE \citep{aydin2025sharpie}, a framework for conducting experiments involving humans and reinforcement learning agents. SHARPIE provides a generic interface that wraps existing Gymnasium environments and supports several human-AI interaction paradigms. While SHARPIE and \mug share similar goals of reducing barriers to human-AI experimentation, they differ in emphasis and capabilities. SHARPIE provides breadth across interaction types (e.g., reward specification, preference elicitation) but does not address the challenges of real-time, low-latency multiplayer interaction, client-side execution, or a complete and configurable experiment flow. \mug is designed specifically to address these issues to provide an experience that is scalable and accessible.} 

\red{Similarly, CrowdPlay \citep{gerstgrasser_crowdplay_2021} is a web-based platform for crowdsourcing human demonstration trajectories in Gymnasium environments. CrowdPlay supports multi-agent games; however, it uses a server-authoritative architecture in which the environment renders frames as JPEG images that are streamed to the browser. No environment logic runs client-side. CrowdPlay provides no latency compensation beyond frame dropping, and its experiment lifecycle is limited (instructions and payment tracking, but no structured scene flow or surveys). Furthermore, the CrowdPlay project was deprecated in 2025 and is no longer maintained.}

\red{HIPPO-Gym \citep{bewley2021hippo} provides a web-based platform for human-in-the-loop RL research, allowing humans to interact with Gymnasium environments through a browser interface. HIPPO-Gym is limited to single-agent settings and supports only specific interaction paradigms (e.g., humans teaching RL agents), without support for multi-human or multi-agent experiments.}

\red{A common limitation across these platforms is their reliance on server-authoritative architectures, where the environment runs on the server and each frame requires a network round trip to exchange actions and state. For single-player settings, this introduces latency proportional to the participant's connection quality. For multiplayer, the problem compounds: all players' inputs must reach the server before the environment can step, and the updated state must be sent back to all clients. At frame rates typical of real-time tasks (e.g., 30--60 frames per second), this can make interactions unusable for participants with moderate latency. \mug addresses this through client-side execution and, for multiplayer, peer-to-peer input exchange with rollback-based synchronization, as described in Section~\ref{sec:interactive-gym}.} \red{Table~\ref{tab:platform_comparison} summarizes the capabilities of existing interactive experiment platforms for RL environments.}

\red{
\begin{table}[htbp]
\centering
\footnotesize
\begin{tabular}{l c c c c c}
\toprule
 & API Compat. & Multi-human & Client-side & Latency comp. & Exp.\ lifecycle \\
\midrule
Overcooked-AI & Custom only & \cmark & \xmark & \xmark & \xmark \\
HIPPO-Gym & Gymnasium & \xmark & \xmark & \xmark & Partial \\
CrowdPlay & Gymnasium & \cmark & \xmark & \xmark & Partial \\
SHARPIE & Gymnasium & \cmark & \xmark & \xmark & Partial \\
\mug & Gymnasium/PettingZoo & \cmark & Pyodide & GGPO & \cmark \\
\bottomrule
\end{tabular}
\caption{\red{Comparison of platforms for interactive experiments with simulation environments. \emph{API Compat.} indicates which environment APIs are natively supported. \emph{Multi-human} indicates simultaneous interaction between multiple human participants. \emph{Client-side} indicates whether the environment can run in the participant's browser. \emph{Latency Comp.} indicates whether the platform provides a mechanism (e.g., rollback netcode) to mask network latency during real-time interaction. \emph{Exp.\ lifecycle} indicates built-in support for the full experiment flow (instructions, matchmaking, surveys, data collection).}}
\label{tab:platform_comparison}
\end{table}
}

\red{As summarized in Tables~\ref{tab:env_comparison} and~\ref{tab:platform_comparison}, existing environment libraries require either non-Python languages for custom environments or lack hardware acceleration, and existing experiment platforms do not jointly support standard simulation environment APIs, real-time multi-human interaction, and client-side execution. \cogrid and \mug are designed to address these gaps.}

\section{\cogrid} \label{sec:cogrid}

All \cogrid~environments are partially observable Markov decision processes (POMDPs), described by the tuple $(\mathcal{X}, \mathcal{A}, \mathcal{O}, \mathcal{T}, \mathcal{R}, \Omega)$. Here, $\mathcal{X}$ is the state space, $\mathcal{A}$ the action space, $\mathcal{O}$ the observation space, $\mathcal{T}:\mathcal{X}\times \mathcal{A}\to \mathcal{X}$ the transition function, $\mathcal{R}:\mathcal{X}\times \mathcal{A}\to \mathbb{R}$ the reward function, and \red{$\Omega:\mathcal{X}\to\mathcal{O}$} the observation function. 

We first review the components of Minigrid that we have utilized in \cogrid, then detail the improvements made and the motivations behind each. 

\paragraph{Minigrid Functionality.}

\red{Minigrid \citep{chevalier-boisvert_minigrid_2023} is a single-agent grid-world library built on the Gymnasium API \citep{towers_gymnasium_2023}. Environments are 2D $n\times m$ grids in which each cell is either empty or occupied by a \pyinline{WorldObj}. The environment tracks the agent's position, inventory, and direction separately from the grid, rather than representing the agent as a \pyinline{WorldObj}---this is the primary reason, at the implementation level, that Minigrid does not support multiple agents. Minigrid provides fixed default observation, action, and reward interfaces: a partial grid view with agent direction and a mission string; a discrete action space for rotation, movement, and object interaction; and sparse rewards upon mission completion. Because Minigrid has no hardware-accelerated backend, simulation speed can become a bottleneck during large-scale training runs. NAVIX \citep{pignatelli2024navix} addresses this by reimplementing Minigrid in JAX for hardware-accelerated simulation, but it retains the single-agent design.}

\red{\paragraph{\cogrid's Dual Backend.} All simulation code in \cogrid operates through a backend-agnostic array namespace (\pyinline{cogrid.backend.xp}) that dispatches to either NumPy or JAX. When the JAX backend is active, environment functions are automatically just-in-time (JIT) compiled, and \pyinline{jax.vmap} can be used for batched execution across many environment instances in parallel. With the NumPy backend, the same code runs without a JAX dependency, which is critical for settings that require Pyodide-compatible libraries (e.g., for client-side execution via WebAssembly in \mug). This single-implementation approach distinguishes \cogrid from libraries such as JaxMARL \citep{flair2024jaxmarl}, which target GPU-accelerated training but cannot run in the browser. A concrete illustration of the benefits of GPU acceleration is provided in Section~\ref{sec:case_study_overcooked} where we demonstrate empirical results in an example environment. In practice, users can swap backends simply by changing the \pyinline{backend} argument in environment construction: \pyinline{cogrid.make("MyEnvironment", backend="numpy")} or \pyinline{cogrid.make("MyEnvironment", backend="jax")}.}

\paragraph{\cogrid Agents.} \cogrid defines agents as their own \pyinline{Agent} object, inheriting from the same \pyinline{GridObj} class that all other environment objects do. This allows the environment to track an arbitrary number of agents within the \pyinline{Grid}.

\paragraph{\cogrid Observations.} Rather than setting defaults for agent observations, we have chosen to customize the observation space individually for each environment. \red{Environments specify features by name in a configuration file. Each \pyinline{Feature} subclass provides a pure function that maps the current state to an observation vector. At initialization, the selected features are composed into a single observation function that concatenates outputs and composes them into a single observation for each agent in the environment.}

\paragraph{\cogrid Actions.} The action space in \cogrid environments is, by default, similar to that of Minigrid. The default action space is discrete and consists of \pyinline{"turn left"}, \pyinline{"turn right"}, \pyinline{"move forward"}, \pyinline{"pickup or drop"}, \pyinline{"toggle"}, and \pyinline{"no-op"}. Picking up and dropping are consolidated into a single action, with the environment logic depending on which cell the agent is facing. The \pyinline{"no-op"} action allows agents the option to do nothing. This has a number of use cases, such as an agent waiting for another. Beyond this default action set, we also provide direct movement actions that eliminate rotations. This allows the \pyinline{"turn left"}, \pyinline{"turn right"}, and \pyinline{"move forward"} actions to be replaced with \pyinline{"move left"}, \pyinline{"move right"}, \pyinline{"move up"} and \pyinline{"move down"}. The agent rotation is then changed to correspond with the direction moved. The motivation for this change, along with the unification of picking up and dropping, was to make the controls more intuitive for a human player if they were controlling an agent in the environment or interacting with other agents. Actions can be added or removed as long as the corresponding environment logic is added in the environment loop. \red{The active action set is selected via the environment configuration.}

\paragraph{\cogrid Rewards.} Similar to the observations in \cogrid, rewards have been modularized into their own class to allow increased flexibility and easy customization. \red{Each \pyinline{Reward} subclass implements a \pyinline{compute} method that receives both the previous and current state together with the actions taken, and returns a per-agent reward array. \cogrid allows for abstractions that will automatically build rewards that are compatible with JAX's JIT-compilation and parallel execution, alleviating complexity for researchers. For example, we provide the \pyinline{InteractionReward} base class that allows users to specify a small set of arguments to define custom rewards around basic interactions with objects. Further detail is provided in Appendix~\ref{sec:overcooked_appendix} and the online documentation.}

\begin{figure}[htbp] 
    \centering
    \begin{subfigure}{0.45\textwidth}
        \centering
        \includegraphics[width=\textwidth]{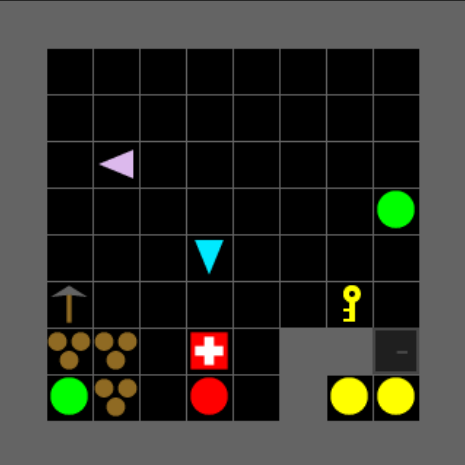}
        \caption{A visualization of the full map from the \cogrid \pyinline{search_rescue} environment. Agents are represented by triangles that point in their current direction. In this environment, two agents must work together to rescue ``victims'' in the environment (green, red, and yellow circles). There is rubble that must be cleared by picking up a pickaxe and doors that can only be opened by picking up a key. Yellow victims can only be rescued when an agent is holding the medical kit (red square with a white cross), and red victims require both agents to rescue them simultaneously.}
    \end{subfigure}\hfill
    \begin{subfigure}{0.45\textwidth}
        \begin{minted}[fontsize=\footnotesize]{python}
class GridAgent(GridObj):
    [...]
    def render(self, tile_img):
        tri_fn = point_in_triangle(
            (0.12, 0.19),
            (0.87, 0.50),
            (0.12, 0.81),
        )

        # Rotate based on agent direction
        tri_fn = rotate_fn(
            tri_fn, 
            cx=0.5, 
            cy=0.5, 
            theta=0.5 * math.pi * self.dir
        )
        fill_coords(
            tile_img, 
            tri_fn, 
            self.color
        )
        \end{minted}
        \caption{\red{Every object has} a \pyinline{render} function, which can be arbitrarily complex. The rendering shown here uses the \red{Minigrid} primitives for drawing a triangle for an agent, rotated based on the agent's direction.}
    \end{subfigure}
    \caption{Visualization of an example environment and sample code to draw the \pyinline{GridAgent} object. The default visualization utilizes the rendering functions developed for Minigrid.}
    \label{fig:vis_example}
\end{figure}

\paragraph{\cogrid Visualization.} \cogrid retains an identical visualization scheme as \red{Minigrid}. In particular, every object in the environment has an associated \pyinline{render} function that adds to an RGB tile rendered at that object's location. It is possible to make these tiles arbitrarily complex; however, most commonly, the images are based on graphical primitives defined through \red{Minigrid's} rendering utilities (e.g., drawing circles, squares, and lines). An example visualization of a simple team-based search and rescue task\footnote{This task is included in \cogrid as an adaptation of the Minimap task developed by \cite{nguyen2023minimap}.} is shown in Figure~\ref{fig:vis_example}.

\section{Multi-User Gymnasium (\mug)} \label{sec:interactive-gym}

While \cogrid provides a library for building grid-based multi-agent environments, \mug serves as a library for conducting interactive experiments with such multi-agent environments. Importantly, \cogrid \emph{is not} a prerequisite for \mug: it is designed to be compatible with any environment that follows the same standardized API and can be customized to work with those that do not. 

We constructed the interface to allow experiments to be designed as a sequence of ``scenes'' (that is, \pyinline{Scene} objects). All scenes represent configurations that a researcher can fill out to create a component of their experiment. Human interaction with simulation environments is achieved through a specific type of scene: a \pyinline{GymScene}. Here, a researcher will specify all the information related to their simulation environment, how users interact with it, and how it is displayed to a participant in their study. This includes mapping keystrokes to actions, providing a mechanism to instantiate simulation environments, specifying AI policies, among much more. Behind the scenes, \mug facilitates client-server communication in order to send relevant data to the client, where a game engine---i.e., Phaser---is running to display the current state of the simulation to the participant. Any required information is then transmitted back from the client to the server to be processed and stored for analysis.\footnote{Data storage may take any form defined by the user through our flexible API. Trajectories are currently automatically exported as sequences of states and actions on the server; however, researchers can define any data storage format or location (e.g., external database).}

\begin{figure}[p!]
    \centering
    \scalebox{0.85}{\input{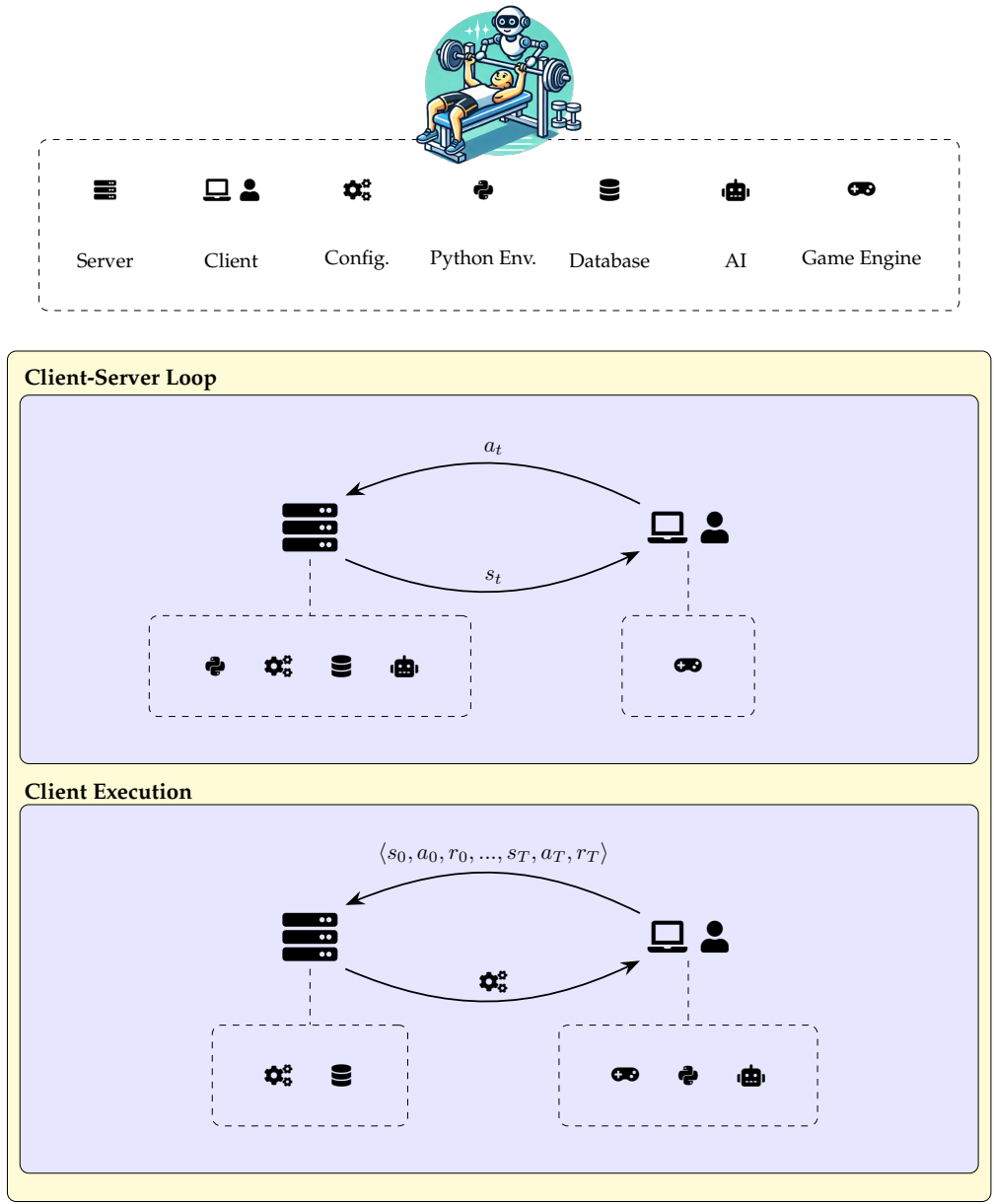}}

    \caption{The two ways in which we run simulation environments in \pyinline{GymScene}s: either through a continuous client-server communication loop or by executing the Python environment code directly in the client's browser. In the former, the server maintains the environment and interactions with it---as well as all AI agents\red{---}and simply accepts data from the user in the form of actions $a_t$, while providing \red{updated} states $s_t$ at every step. In the second case, the server provides all data necessary---through the \mug configuration---for the client to execute the Python code in their browser.}
    \label{fig:environment_protocol_ig}

\end{figure}

An important specification that can be made is \emph{where} the environment is executed. A typical approach to allow Python-based applications to interact with web-based clients would be to have a continuous communication loop where all Python code is executed on the server and updated information is passed to the client on each ``tick.'' \red{This is the approach of both SHARPIE and HIPPO-Gym, and it works in cases where participants have very low latency connections to the server or the tick rate of the environment is low.} In cases where this communication is frequent (e.g., a simulation environment updating many times a second), network latency can make this approach ineffective. The novel approach we have developed in \mug is to execute Python-based environments directly in client browsers \red{using Pyodide,} which eliminates the need for this communication and allows researchers to serve their experiments to participants who may have high-latency connections. The distinction and general construction of \pyinline{GymScene}s are shown in Figure~\ref{fig:environment_protocol_ig}.

\red{Client-side execution removes the latency problem for single-player experiments. However, multiplayer introduces a new challenge. If each client runs its own copy of the environment, their states must remain synchronized. The common approach in existing work, where each client waits for all players' inputs before stepping, reintroduces the same latency that client-side execution was designed to avoid. To address this, \mug implements a form of GGPO~\citep{cannon2019ggpo} rollback netcode, a technique originally developed for latency-sensitive fighting games. Each client runs the environment locally with a shared random seed, applies its own input immediately, and predicts the other remote players' input (by default, repeating their last known action). When another player's input arrives that differs from the prediction, the client rolls back to the last confirmed state and replays the intervening frames with the correct inputs. This process is illustrated in Figure~\ref{fig:ggpo_rollback}. Critically, to enable this functionality the environments must be deterministic given a random seed and must implement state retrieval and setting (\pyinline{get_state()} and \pyinline{set_state()} methods for MUG compatibility).}

\red{
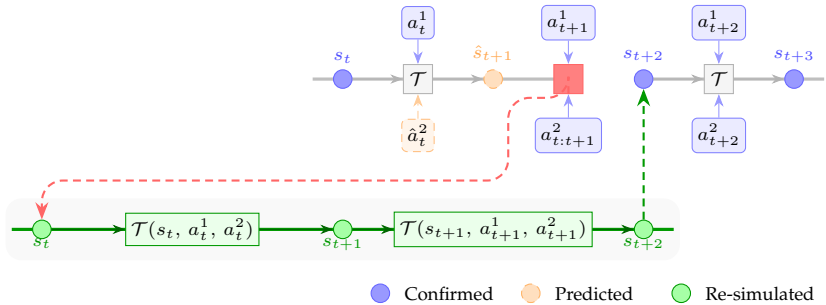
\begin{figure}[htbp]
    \centering
    \begin{tikzpicture}[
    >=Stealth,
    timeline/.style={very thick, gray!50},
    stateC/.style={circle, fill=blue!40, draw=blue!60, minimum size=7pt, inner sep=0pt},
    stateP/.style={circle, fill=orange!25, draw=orange!60, densely dashed,
                   minimum size=7pt, inner sep=0pt},
    stateR/.style={circle, fill=green!35, draw=green!60!black, minimum size=7pt, inner sep=0pt},
    trans/.style={rectangle, draw=gray!60, fill=gray!8, minimum size=11pt,
                  inner sep=1pt, font=\scriptsize},
    act/.style={rounded corners=2pt, draw, font=\scriptsize, inner sep=2pt},
    lbl/.style={font=\scriptsize},
    statelbl/.style={font=\scriptsize, inner sep=0pt},
]

\def\dx{2.0}

% ====== MAIN TIMELINE (two separate segments — break after red box) ======
\draw[timeline] (-0.4,0) -- ({1.5*\dx + 0.15},0);
\draw[timeline] ({2*\dx},0) -- ({3*\dx+0.4},0);

% Tick marks at state positions
\foreach \i in {0,...,3} {
    \draw[gray!40] ({\i*\dx},-0.1) -- ({\i*\dx},0.1);
}

% State dots on main timeline (labels above)
\node[stateC] (s0) at (0,0) {};
\node[statelbl, blue!70, above=6pt] at (s0) {$s_t$};

\node[stateP] (s1) at (\dx,0) {};
\node[statelbl, orange!70, above=6pt] at (s1) {$\hat{s}_{t{+}1}$};

\node[stateC] (s2) at ({2*\dx},0) {};
\node[statelbl, blue!70, above=6pt] at (s2) {$s_{t{+}2}$};

\node[stateC] (s3) at ({3*\dx},0) {};
\node[statelbl, blue!70, above=6pt] at (s3) {$s_{t{+}3}$};

% ====== TRANSITION BOXES (at midpoints on timeline) ======
\node[trans] (T0) at ({0.5*\dx}, 0) {$\mathcal{T}$};
\node[rectangle, draw=red!60, fill=red!50, minimum size=11pt, inner sep=1pt] (T1) at ({1.5*\dx}, 0) {};
\node[trans] (T2) at ({2.5*\dx}, 0) {$\mathcal{T}$};

% ====== CLIENT 1 ACTIONS (above, at midpoints between states) ======
% \node[font=\small\bfseries, anchor=east] at (-0.6, 0.75) {Client 1};
\node[act, draw=blue!50, fill=blue!8] (a10) at ({0.5*\dx}, 0.75) {$a^1_t$};
\node[act, draw=blue!50, fill=blue!8] (a11) at ({1.5*\dx}, 0.75) {$a^1_{t{+}1}$};
\node[act, draw=blue!50, fill=blue!8] (a12) at ({2.5*\dx}, 0.75) {$a^1_{t{+}2}$};

% Arrows: Client 1 actions → T boxes
\draw[->, blue!40] (a10.south) -- (T0.north);
\draw[->, blue!40] (a11.south) -- (T1.north);
\draw[->, blue!40] (a12.south) -- (T2.north);

% ====== CLIENT 2 ACTIONS (below, at midpoints between states) ======
% \node[font=\small\bfseries, anchor=east] at (-0.6, -0.75) {Client 2};

% Predicted at t (Client 2's input not received → predicted)
\node[act, draw=orange!50, densely dashed, fill=orange!8] (a20) at ({0.5*\dx},-0.75) {$\hat{a}^2_t$};

% Late arrival bundle (actual inputs for t and t+1 arrive together)
\node[act, draw=blue!50, fill=blue!8] (late) at ({1.5*\dx},-0.75) {$a^2_{t{:}t{+}1}$};

% On-time at t+2
\node[act, draw=blue!50, fill=blue!8] (a22) at ({2.5*\dx},-0.75) {$a^2_{t{+}2}$};

% Arrows: Client 2 actions → T boxes
\draw[->, orange!40, densely dashed] (a20.north) -- (T0.south);
\draw[->, blue!40] (late.north) -- (T1.south);
\draw[->, blue!40] (a22.north) -- (T2.south);

% Arrows: state → T boxes (state is input to transition)
\draw[->, gray!60] (s0.east) -- (T0.west);
\draw[->, gray!60] (s2.east) -- (T2.west);

% Arrows: T boxes → next state (skip T1 — invalidated, triggers rollback)
\draw[->, gray!60] (T0.east) -- (s1.west);
\draw[->, gray!60] (T2.east) -- (s3.west);

% ====== ROLLBACK TIMELINE (wider spacing, r2 aligned below s2) ======
\def\rdx{4.0}
\draw[very thick, green!60!black] ({2*\dx - 2*\rdx - 0.4},-2.0) -- ({2*\dx+0.4},-2.0);

% Rollback ticks
\draw[green!40!black] ({2*\dx - 2*\rdx},-2.1) -- ({2*\dx - 2*\rdx},-1.9);
\draw[green!40!black] ({2*\dx - \rdx},-2.1) -- ({2*\dx - \rdx},-1.9);
\draw[green!40!black] ({2*\dx},-2.1) -- ({2*\dx},-1.9);

% Rollback state dots (labels below)
\node[stateR] (r0) at ({2*\dx - 2*\rdx},-2.0) {};
\node[statelbl, green!60!black, below=3pt] at (r0) {$s_t$};

\node[stateR] (r1) at ({2*\dx - \rdx},-2.0) {};
\node[statelbl, green!60!black, below=3pt] at (r1) {$s_{t{+}1}$};

\node[stateR] (r2) at ({2*\dx},-2.0) {};
\node[statelbl, green!60!black, below=3pt] at (r2) {$s_{t{+}2}$};

% Rollback transition boxes (on the line at midpoints)
\node[trans, draw=green!60!black, fill=green!8, inner sep=2pt] (rT0) at ({2*\dx - 1.5*\rdx}, -2.0)
    {$\mathcal{T}(s_t,\, a^1_t,\, a^2_t)$};
\node[trans, draw=green!60!black, fill=green!8, inner sep=2pt] (rT1) at ({2*\dx - 0.5*\rdx}, -2.0)
    {$\mathcal{T}(s_{t{+}1},\, a^1_{t{+}1},\, a^2_{t{+}1})$};

% Arrows: rollback state → T boxes (state is input to transition)
\draw[->, green!40!black] (r0.east) -- (rT0.west);
\draw[->, green!40!black] (r1.east) -- (rT1.west);

% Arrows: rollback T boxes → next rollback state
\draw[->, green!40!black] (rT0.east) -- (r1.west);
\draw[->, green!40!black] (rT1.east) -- (r2.west);

% Subtle background
\begin{pgfonlayer}{background}
    \node[draw=none, fill=gray!5, rounded corners=6pt,
          inner xsep=10pt, inner ysep=8pt,
          fit=(r0)(r2)] {};
\end{pgfonlayer}

% ====== CONNECTIONS ======
% Trigger: red T box → rewind to checkpoint (routed left of late action box)
\draw[->, red!60, thick, densely dashed, rounded corners=8pt]
    (T1.south) -- ({1.5*\dx},-0.25) -- ({1.1*\dx},-0.25)
    -- ({1.1*\dx},-1.35) -- ({2*\dx - 2*\rdx},-1.35) -- (r0.north);

% Merge: corrected result → main timeline (straight up)
\draw[->, green!60!black, thick, densely dashed]
    (r2.north) -- (s2.south);

% ====== LEGEND ======
\begin{scope}[yshift=-2.85cm, xshift=0.5cm]
    \node[stateC] at (0,0) {};
    \node[lbl, anchor=west] at (0.2, 0) {Confirmed};
    \node[stateP] at (2.0,0) {};
    \node[lbl, anchor=west] at (2.2, 0) {Predicted};
    \node[stateR] at (4.0,0) {};
    \node[lbl, anchor=west] at (4.2, 0) {Re-simulated};
\end{scope}

\end{tikzpicture}
    \caption{\red{Illustration of GGPO rollback netcode in a two-player environment. The main timeline shows the simulation state at each tick; each transition $\mathcal{T}(s_t, a^1_t, a^2_t) \to s_{t+1}$ requires both players' actions. Client~1's actions ($a^1$) arrive on time at every tick. Client~2's action at $t{+}1$ is delayed, so Client~1 predicts it ($\hat{a}^2$), producing a speculative state. When the delayed input arrives as a bundle $a^2_{t{:}t+1}$ at $t{+}1$ (green), the client rolls back to the last confirmed state $s_t$ and re-simulates with the correct actions (green timeline), merging the corrected state back into the main timeline. Rollback occurs without rendering re-simulated frames and it takes place between rendered ticks, causing minimal visual disruption.}}
    \label{fig:ggpo_rollback}
\end{figure}
}

\begin{figure}[htbp]    

    \centering
    \input{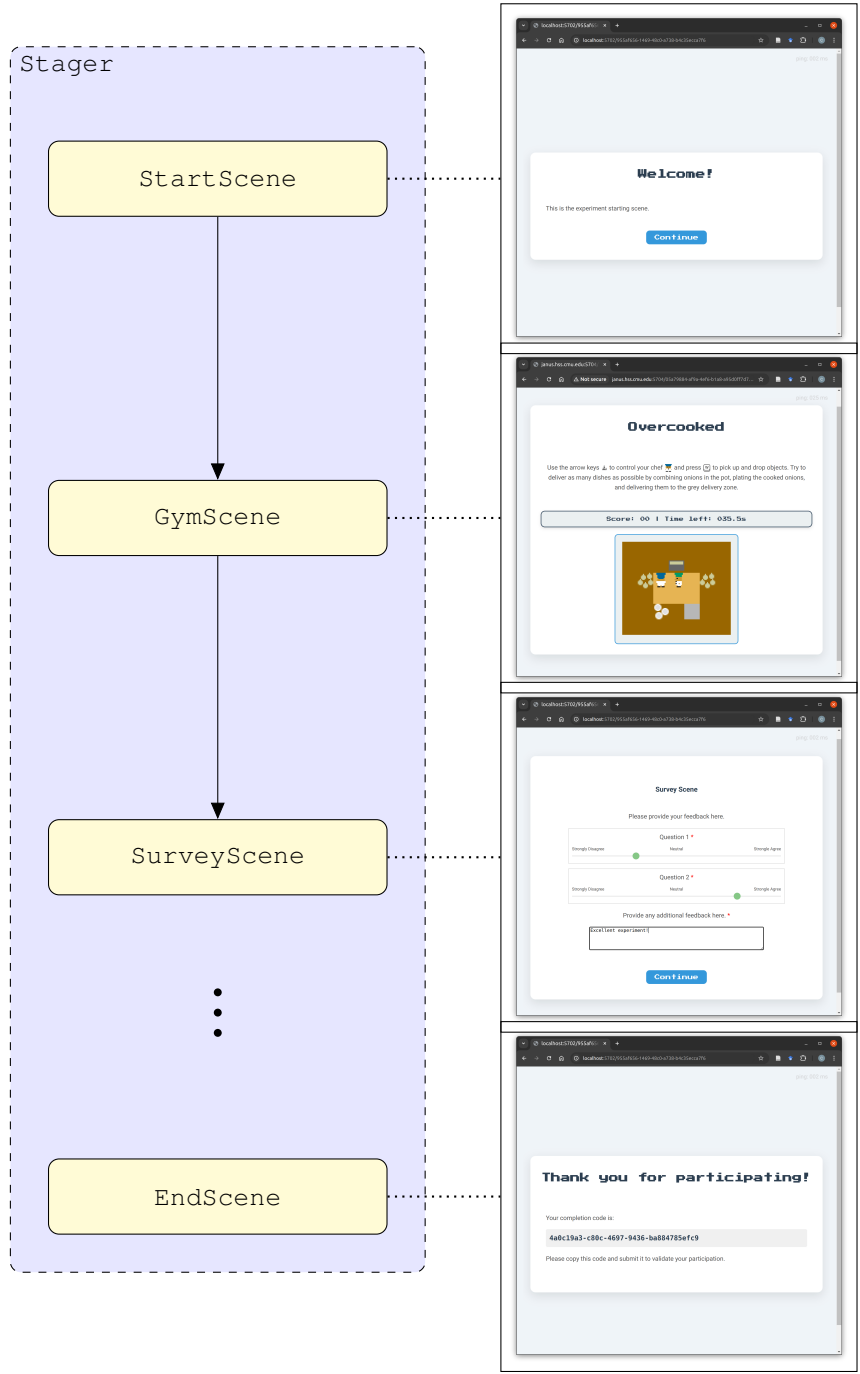}
    \caption{\mug experiment flow using a \pyinline{Stager} and sequence of \pyinline{Scene}s. The \pyinline{Stager} defines the flow of the experiment and also has the capabilities to manipulate the order or assignment at the participant level for experiments with multiple conditions.}
    \label{fig:ig_stager_flow}
  
\end{figure}

Beyond the ability to define interaction with simulation environments in a \pyinline{GymScene}, we have created a collection of alternative scenes that have distinct functionalities and customization. This enables researchers to modularize their experiments and integrate customized components or pages that do not require interaction with simulation environments. For example, static informational pages can be created alongside surveys or alternative forms of interaction. When constructing an experiment, all scenes are collected into a \pyinline{Stager} that defines their order---and any manipulation (e.g., randomization)---and the flow of the experiment. This is illustrated in Figure~\ref{fig:ig_stager_flow}.

\paragraph{Data Collection.} In \mug, data collection is handled through hooks that are called throughout the environment loop. Users can define \pyinline{GameCallback}s that provide hooks at each key environment event, such as episodes starting and ending, each environment step, players joining and leaving, among others. This provides functionality for data to be recorded (e.g., actions and environment state at each environment step) and saved either locally or to an external database. By default, all information returned in the \pyinline{infos} dictionary from the environment is logged at every step and stored on the disk. If a use case requires logging customized data to an external database, for example, a user would write their explicit logic into a \pyinline{GameCallback} that logs data to that source at a desired frequency.

\red{\paragraph{Additional Functionality.} Beyond the core experiment loop, \mug provides several utilities common to online behavioral research. Completion codes can be generated automatically for integration with recruitment platforms such as Prolific or Amazon Mechanical Turk. Scene order can be randomized or counterbalanced at the participant level through the \pyinline{Stager}. Participants can also be filtered on the basis of browser compatibility, device type, or measured network latency before they enter the experiment. Similarly, for multi-player experiments, we provide an interface for creating custom matchmaking logic. This can be configured based on a number of attributes, from task performance to peer-to-peer latency. Full descriptions of these features and their usage are available in the online documentation.}

\section{Proof of Concept: Case Studies} \label{sec:case_studies}

To illustrate the functionality of \cogrid and \mug, we provide several demonstrations of experimental use cases: pure simulation, human-AI interaction, and human-human interaction. Any extension or variation is possible within these regimes, including multiple humans with any number of AI agents. These case studies are not intended as definitive psychological experiments, but as proof-of-concept demonstrations showing how \cogrid and \mug come together to offer (1) environment customization, (2) integration of trained agents into human-AI studies, and (3) end-to-end browser-based experiment deployment with Python environments and variable numbers of human participants. We provide methodological details to illustrate how researchers can adapt these pipelines for their own questions.

For transparency, we also report the experimental design elements used in these demonstrations (e.g., recruitment, task structure, compensation). These details are provided not as empirical validation, but to illustrate the methodological templates that researchers can adapt for their own studies. All experiments conducted here use server-side execution for Python environments but are also compatible with browser-based execution.\footnote{See the online documentation at \url{multi-user-gymnasium.readthedocs.io} for examples of browser-based versions of these experiments.}

The first use case is solely in the \cogrid environment, demonstrating the ease of defining a new multi-agent environment and training a reinforcement learning agent in that environment. We recreated the popular Overcooked environment originally developed by \citet{carroll_utility_2020}. In it, we trained a reinforcement learning policy with the Proximal Policy Optimization \citep[PPO;][]{schulman_proximal_2017} algorithm to complete the task. This process is detailed in Appendix~\ref{sec:training_rl_cogrid}, and the efficacy of the agent and environment is demonstrated in the following experiments.

Next, we demonstrate how \mug can be utilized for interactive human studies, both with and without integration with \cogrid. For the former, we demonstrate how agents trained in the \cogrid Overcooked environment can be integrated into a human-AI experiment. For the latter, we use an external environment that is not grid-based, Slime Volleyball \citep{slimevolleygym}.

Lastly, we conducted studies with humans alone. In addition to human-AI experimentation, \mug enables researchers to run experiments where all actors are controlled by human players. Use cases can include data collection for imitation learning or simply collecting human data for the analysis of human-human interactions. 

All studies with human participants were conducted with approval from Carnegie Mellon's Institutional Review Board, and participants were sourced from Amazon Mechanical Turk. For approximately 15 minutes of total experiment time, participants were compensated with a \$1.50 base payment and an additional \$1.50 possible bonus payment based on task performance: \$0.05 per point scored in Slime Volleyball and \$0.03 per dish delivered in Overcooked, capped at \$1.50.

\subsection{\cogrid Overcooked} \label{sec:case_study_overcooked}

The first case study illustrates how \cogrid supports flexible grid-based environment design and customization, and how \mug then enables us to use such environments for human-human and human-AI web-based experiments. 

\begin{figure}
    \centering
        \includegraphics[width=0.75\textwidth]{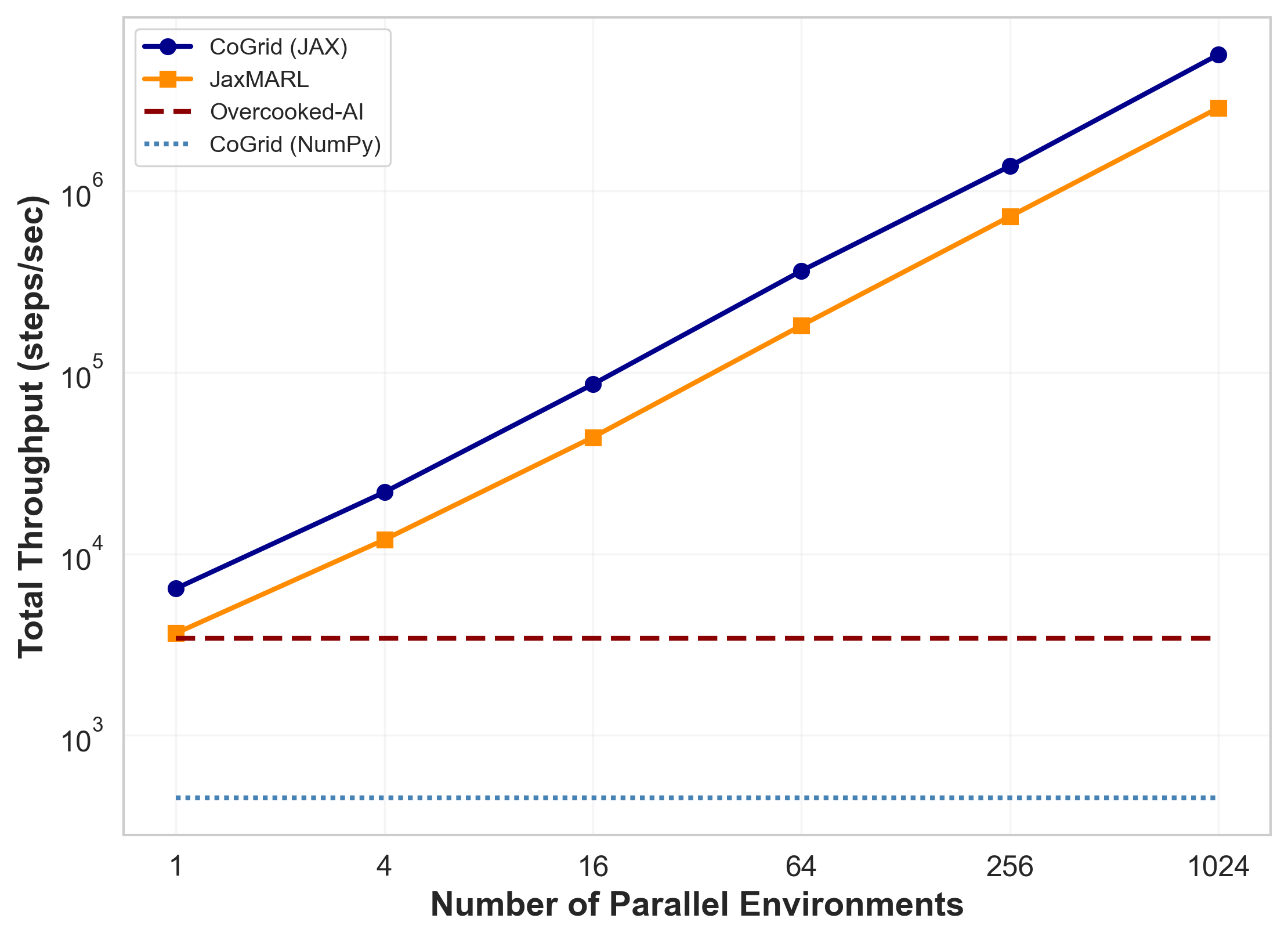} % Replace with your image path
        \caption{\red{Environment throughput in the CoGrid Overcooked environment, comparing to the original Overcooked-AI \citep{carroll_utility_2020} and JaxMARL~\citep{flair2024jaxmarl} implementations. The former has a constant rate of roughly 3,400 steps per second, while the latter scales from roughly 4,300 with a single instance to 2.9 million with 1,024 parallel instances. \cogrid's JAX backend is competitive: scaling from 4,500 steps per second with a single instance to 5.6 million with 1,024 instances, leading to a 1.9x throughput improvement. \cogrid's NumPy backend is by far the slowest at roughly 450 steps per second; however, it offers a a mode that enables browser-based execution via WebAssembly that is entirely absent with JaxMARL. Hardware accelerated execution was run on a single NVIDIA GeForce RTX 3090.}} 
        \label{fig:overcooked_sps}
\end{figure}

\red{We first demonstrate the benefits of the dual-backend approach to \cogrid. Hardware acceleration in JAX is a result of being able to parallelize operations, enabling running many instances of the environment at the same time. To illustrate the magnitude of the speedup, we show the efficiency gain when increasing the number of parallel instances from 1 to 1,024 in Figure~\ref{fig:overcooked_sps}. Hardware acceleration offers a drastic increase in throughput, and \cogrid demonstrates competitive performance with JaxMARL.}

\begin{figure}[htbp]    
    \centering
    \begin{subfigure}{0.35\textwidth}
        \includegraphics[width=\textwidth]{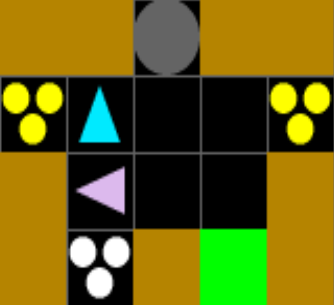} % Replace with your image path
        \caption{The \cogrid Overcooked Cramped Room implementation. The two chefs are represented by triangles, three yellow circles represent a stack of onions, three white circles a stack of plates, the grey circle a pot, and the green square the delivery zone.} 
        \label{fig:cogrid_overcooked}
    \end{subfigure}
    \hspace{.05cm} % adjust the spacing

    \begin{subfigure}{0.45\textwidth}
        \centering
        \includegraphics[width=\textwidth]{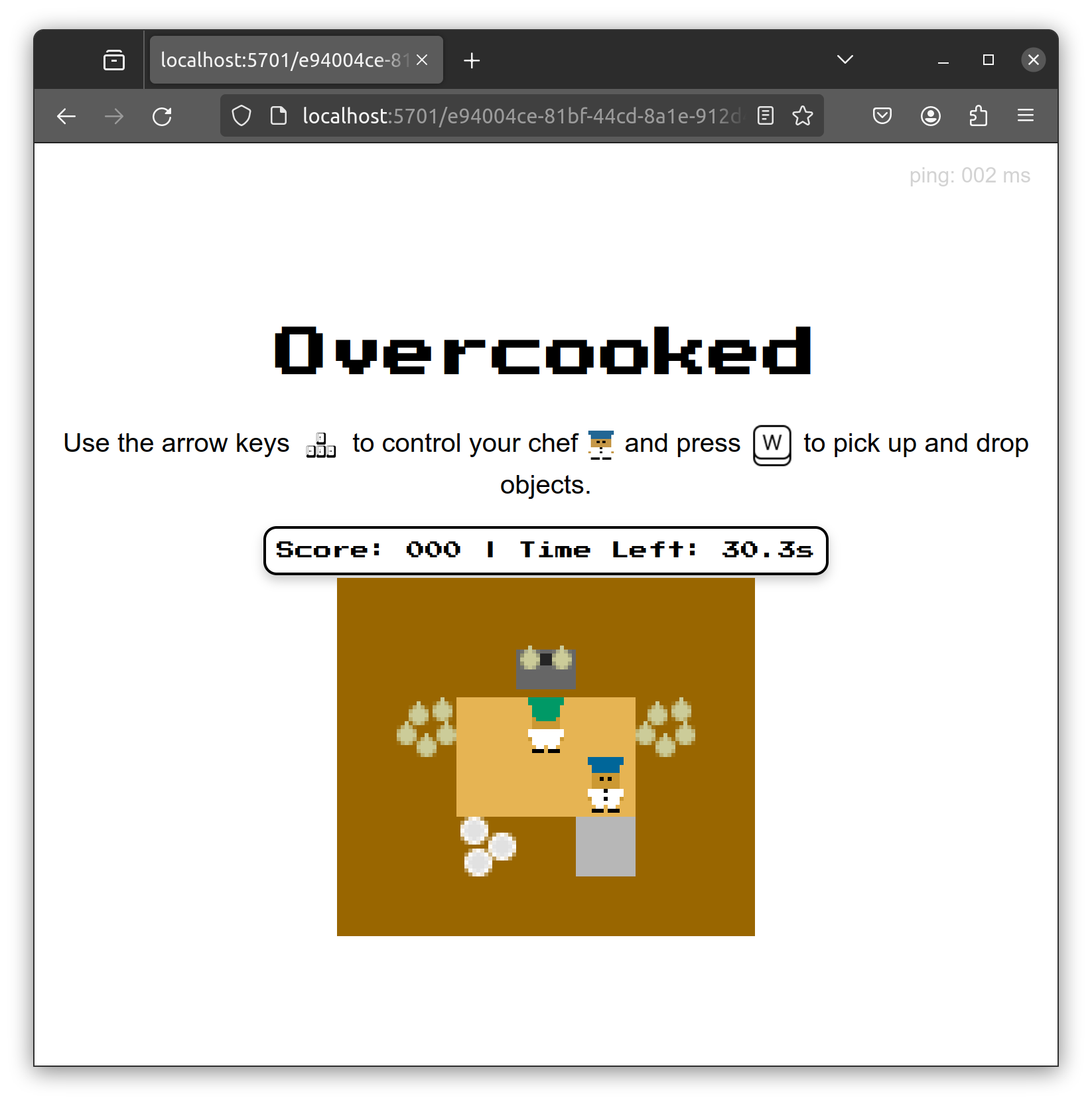} % Replace with your image path
        \caption{A view of the web page that participants saw after they started the task in their browser. The graphics used to render the game are taken from \citet{carroll_utility_2020} and their original open-source implementation of Overcooked-AI. The instructions and icons are part of a customized HTML file that is passed to the \mug configuration. The score and time left display are part of a dynamic block, updated with a Python function that returns the values to populate the display.}
        \label{fig:overcooked_gameplay}
    \end{subfigure}\hfill    
    \caption{The Overcooked visualization from \cogrid's native rendering and visualized in the browser with \mug. The latter uses the assets originally used by \citet{carroll_utility_2020}.}
    \label{fig:overcooked_vis}
\end{figure}

\red{As} an illustration of \cogrid, we implemented a replica of the Cramped Room layout in the Overcooked-AI environment originally developed by \citet{carroll_utility_2020}. In this cooperative task, two players coordinate to prepare and deliver onion soup, which requires coordinated sequential actions to combine ingredients, cook, and plate. Figure~\ref{fig:cogrid_overcooked} shows the layout rendered using the \red{basic rendering functions}. This case study demonstrates how \cogrid enables researchers to reproduce complex cooperative environments with minimal effort. The full implementation and environment details are described in the online documentation.

\paragraph{Training an Overcooked Agent.} We then trained a reinforcement learning agent in the \cogrid Overcooked environment using RLlib \citep{liang2018rllib} with the Proximal Policy Optimization \citep[PPO;][]{schulman_proximal_2017} algorithm. Training used self-play,\footnote{Self-play is a training paradigm in multi-agent reinforcement learning where one policy controls all agents and interacts with itself.} and the trained agent achieved an average of 7.5 dishes delivered per episode. \red{The training that produced the policy in the following experimentation was run using the NumPy backend of \cogrid, rather than the JAX-accelerated variant. An illustration of training with the JAX backend, which takes a small fraction of the time, is provided in the documentation and source repository.}

\paragraph{Human Experiments via \mug.} Finally, we deployed the environment in a browser-based study using \mug. \mug allows for the specification of all required settings for experimentation, including mapping keyboard buttons to actions, creating landing pages and surveys, and specifying the environment display format. The experience for participants in the study can be seen in Figure~\ref{fig:overcooked_gameplay}. These human experiments demonstrate how \mug can transform a \cogrid environment into an interactive online experiment.

We conducted Human-Human and Human-AI studies through separate solicitations. We recruited 29 participants for the Human-AI study and 40 participants (20 pairs) for the Human-Human study through Amazon Mechanical Turk. Each participant completed 20 episodes of the Overcooked task with either the same human partner or a fixed reinforcement learning partner policy, as previously described. Participants first completed a consent form, then proceeded to the \mug landing page with task instructions and a start button. 

Participants in the Human-Human study were sent to a waiting room (minimum duration of 5 seconds) until they were paired with another participant. Human-AI study participants saw a simulated waiting room (randomized duration of 5-25 seconds). After completing the 20 episodes, all participants were redirected to a post-experiment questionnaire. The questionnaire asked a series of questions about relative contributions and behavior, including whether or not the participant thought their partner was a human or an AI. To ensure data quality, we applied exclusion criteria (participants with inactivity at or above 97.5\% or those who left the webpage for 90\% of an episode), leaving 23 participants in our Human-AI study and 36 (18 pairs) in the Human-Human study. 

\begin{figure}
    \centering
    \includegraphics[width=0.75\textwidth]{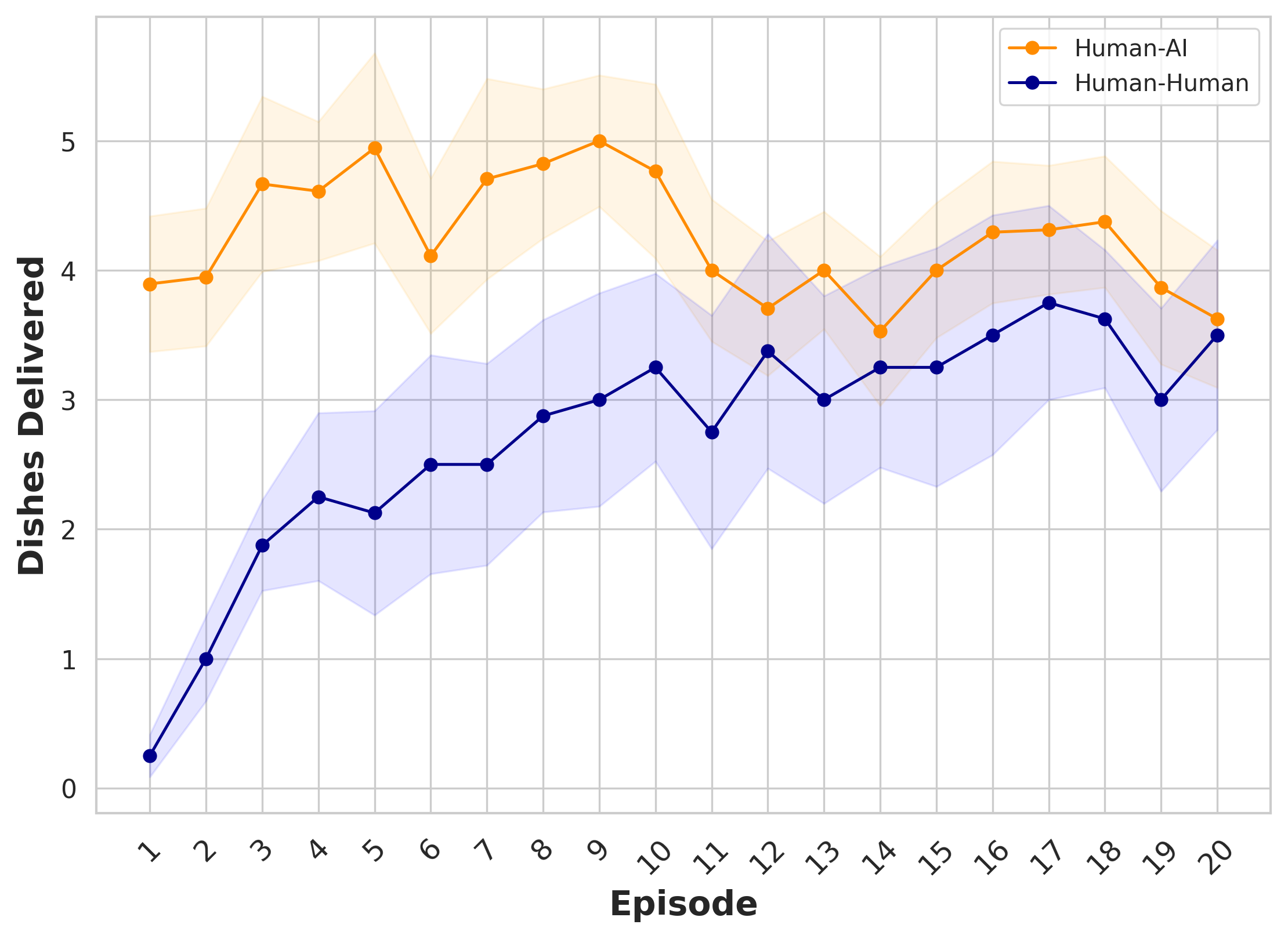}
    \caption{Performance across episodes in both Overcooked studies. The reinforcement learning agent used, when playing with itself with a frame-skip of five, delivered an average of 7.5 dishes per episode (not displayed in the figure). \red{Shaded regions represent standard error.}}
    \label{fig:overcooked_dishes_delivered}
\end{figure}

\paragraph{Results.} These results illustrate the types of analyses enabled by data collected through \mug. The performance results for the Human-Human and Human-AI pairs for each episode are shown in Figure~\ref{fig:overcooked_dishes_delivered}. The relative performance of AI-AI pairs and Human-AI pairs is consistent with the original results of \citet{carroll_utility_2020}, with AI-AI pairs (roughly 7.5 dishes per episode) \red{substantially} outperforming human-AI pairs. Human-AI pairs showed relatively stable performance across episodes\red{, with no statistically significant change in performance over time. In contrast, human-human pairs improved steadily. The stability of human-AI performance is consistent with the findings of \citet{carroll_utility_2020}: the reinforcement learning agent can complete most of the task independently, so although humans likely continue to learn, their improvement is masked by the agent's high baseline capability. In contrast, human-human pairs must develop shared strategies from scratch, and the upward trend reflects this gradual coordination process.}

% This exemplifies the importance of the different conditions and important distinctions between human-AI collaboration and human-human collaboration --- these tools are important because 

% In the Human-AI study, it is likely that human learning is still exhibited; however, the score remains high because the reinforcement learning agent is capable of completing much of the task on its own. This is in line with the observations by \citet{carroll_utility_2020} that reinforcement learning agents are able to accomplish much of the task independently in this environment layout.

\begin{figure}
    \centering
    \includegraphics[width=0.75\textwidth]{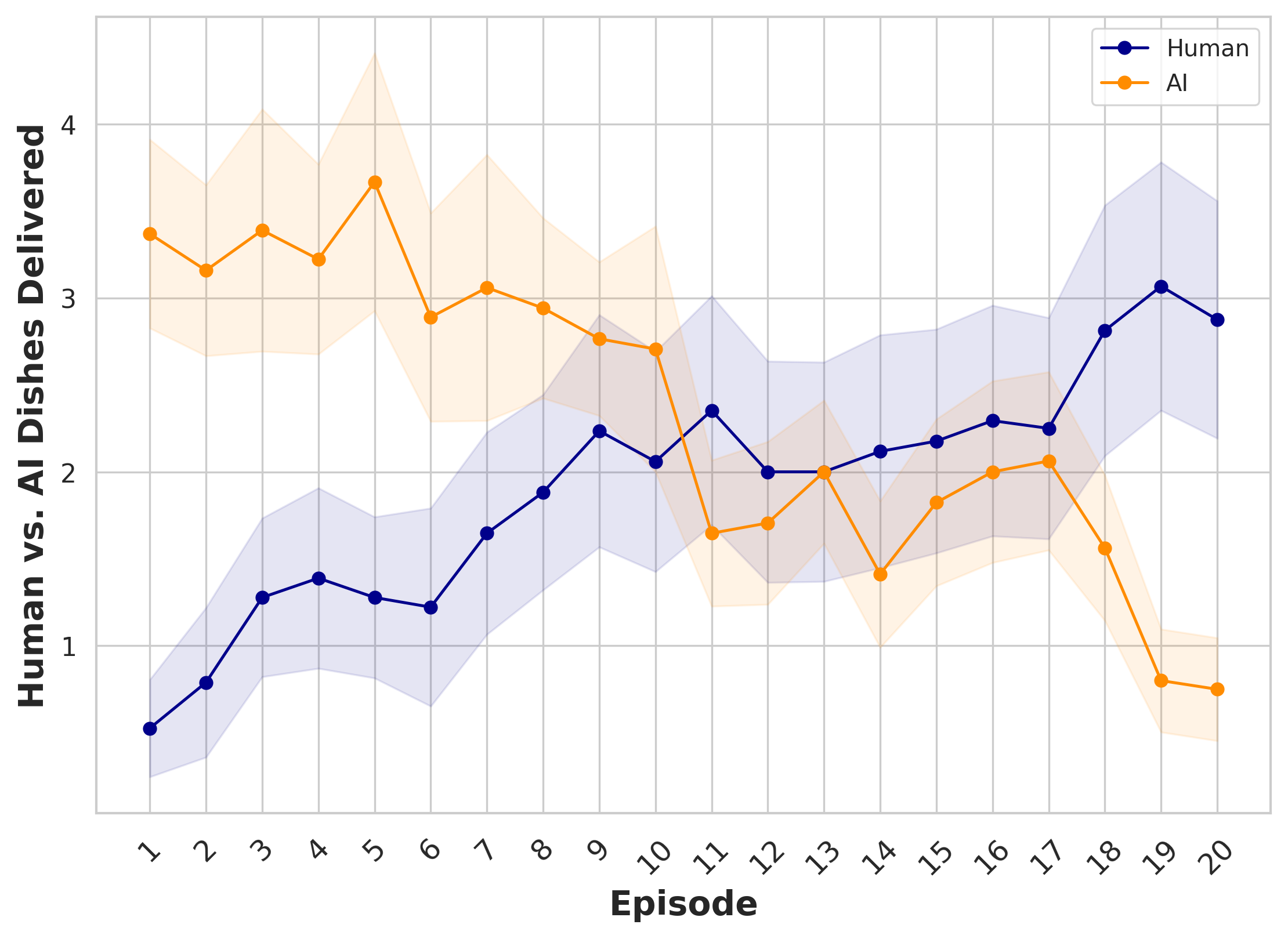}
    \caption{The relative contributions in the Human-AI study, the number of dishes delivered by the human and AI separately.}
    \label{fig:overcooked_contributions}
\end{figure}

We also show the relative contributions of humans and AI in the Human-AI study, as shown in Figure~\ref{fig:overcooked_contributions}. We measured the number of instances where either the human or AI delivered the dish to the delivery zone. \red{Over time, humans adapt their behavior and begin contributing a larger share of deliveries. In interacting with AI, humans appear to be more inclined to complete the dish delivery, rather than taking on the role of putting onions in the pot, even though it results in no observable change in overall task performance.} The results in Figure~\ref{fig:overcooked_contributions} also demonstrate the types of behavioral phenomena that can be captured through studies in human-AI interaction. This kind of behavioral trend---constant performance but varying contributions---exemplifies psychological or cognitive questions in human-AI interaction that can be pursued with
\mug and \cogrid. 

\subsection{Slime Volleyball}  \label{sec:case_study_slimevb}

Next, to complement the Overcooked case study, we demonstrate \mug's compatibility with environments not based on \cogrid. We integrated Slime Volleyball~\citep{slimevolleygym} as a recreation of a classic 2D video game. In the game, \red{players} move laterally and jump to keep a ball \red{off the ground on} their side of the court. A game is won when the ball lands on the court of the other player, similar to volleyball. 

We again trained a reinforcement learning agent using the PPO algorithm. The environment was also fully ported through \mug. Full training and implementation details are provided in Appendix~\ref{sec:slime_vb_training_appendix}. Figure~\ref{fig:slime_interactive_example} shows the original game visualization, as well as the \mug participant view. 

\begin{figure}[htbp]    
    \centering
    \begin{subfigure}{0.45\textwidth}
        \includegraphics[width=\textwidth]{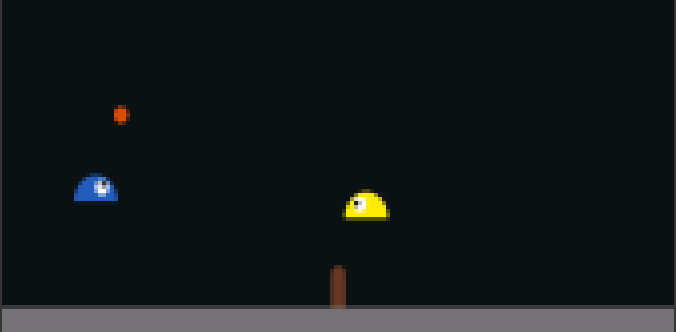} % 
        \caption{The original Slime Volleyball visualization from \citet{slimevolleygym}.}  
    \end{subfigure} \\
    \begin{subfigure}{0.75\textwidth}
        \centering
        \includegraphics[width=\textwidth]{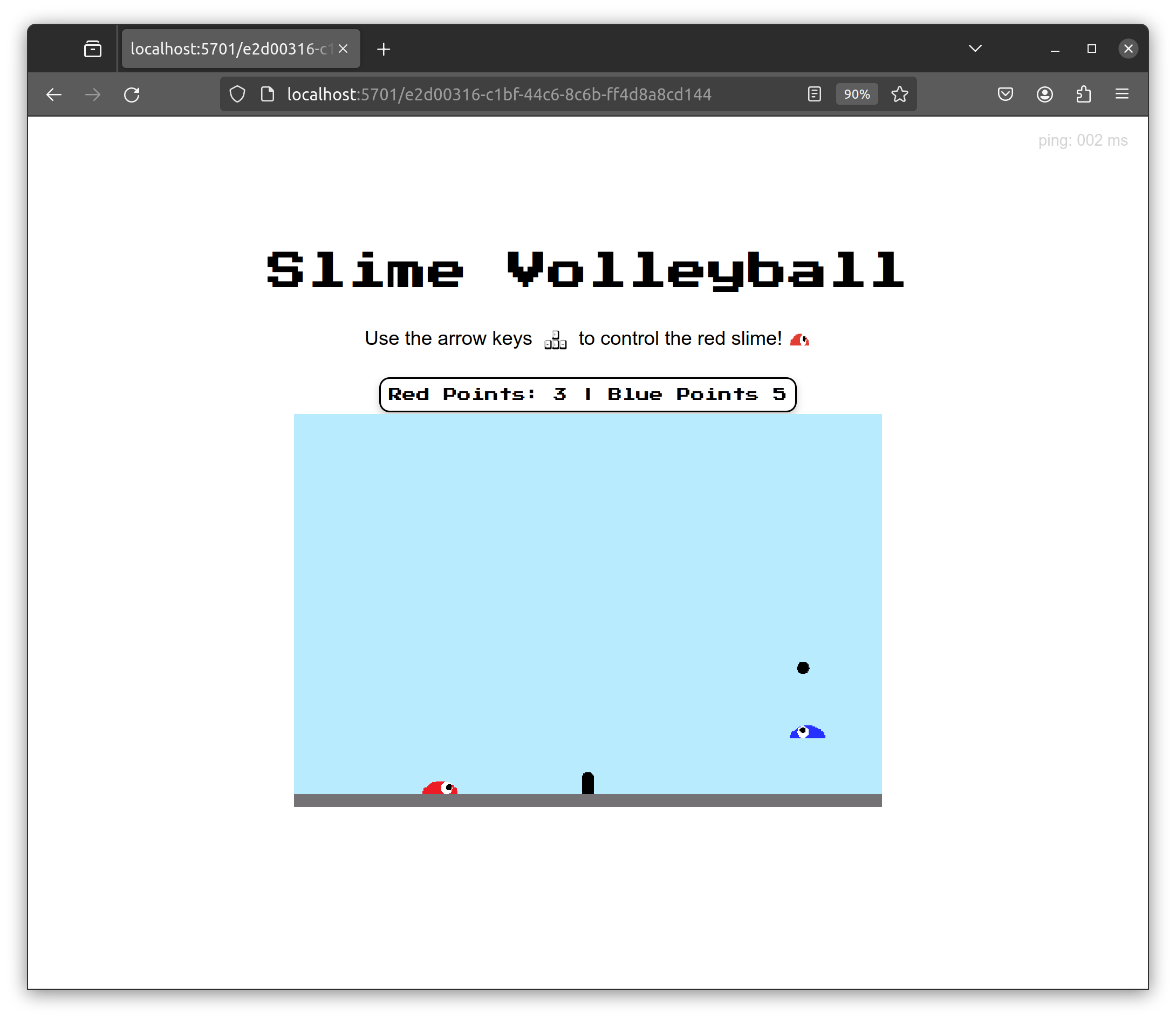} % Replace with your image path
        \caption{Participants' view of Slime Volleyball during the study, as implemented in \mug.}
    \end{subfigure}\hfill    
    \caption{The original and \mug Slime Volleyball interfaces.}
    \label{fig:slime_interactive_example}
\end{figure}

The experimental procedure mirrored that of the Overcooked study---the flow from the consent form, to instructions, waiting room, gameplay, and then the follow-up questionnaire remains unchanged. Participants completed 30 episodes of the task, each consisting of a single point. We recruited 21 participants for the Human-AI condition and 22 participants (11 pairs) for the Human-Human condition from Amazon Mechanical Turk. As before, we excluded participants for data quality. Participants were excluded in instances where they exceeded a maximum RTT,\footnote{The RTT is the ``round trip time'' of a message to go from the server to the participant and back. Given the fast-paced nature of the game, poor connections result in negative experiences.} navigated away from the web page for 90\% of the experiment, or were inactive (97.5\% no-op actions). This resulted in dropping one participant from the Human-AI study.  

\begin{figure}
    \centering
    \includegraphics[width=0.75\textwidth]{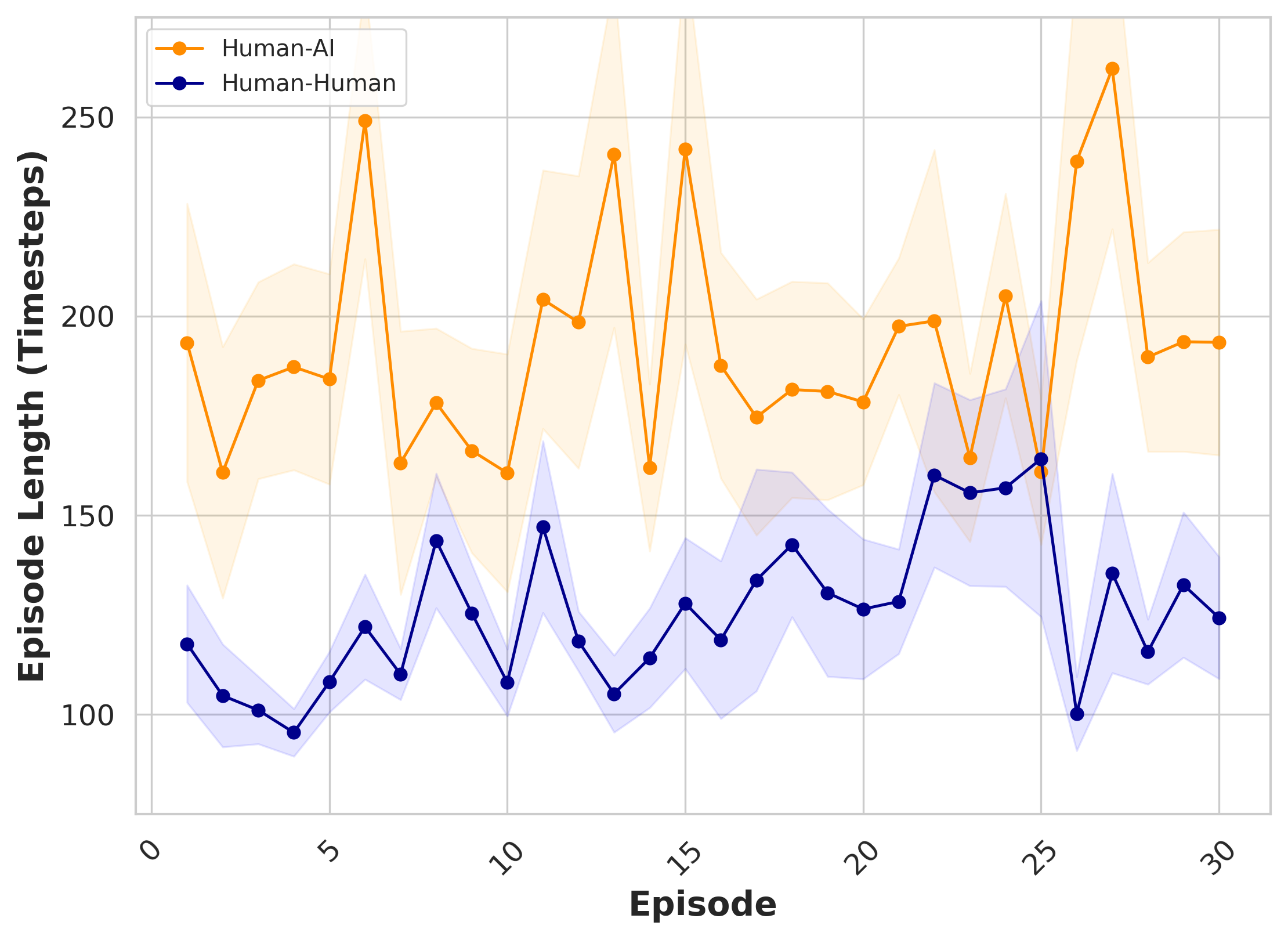}
    \caption{The average episode length over time in the human-AI and human-human studies of Slime Volleyball. The average episode length is an effective proxy for skill because it measures how long the ball is kept in the air, which requires competency. \red{Shaded regions represent standard error.}}
    \label{fig:slime_vb_ep_len}
\end{figure}

\begin{figure}
    \centering
    \includegraphics[width=0.75\textwidth]{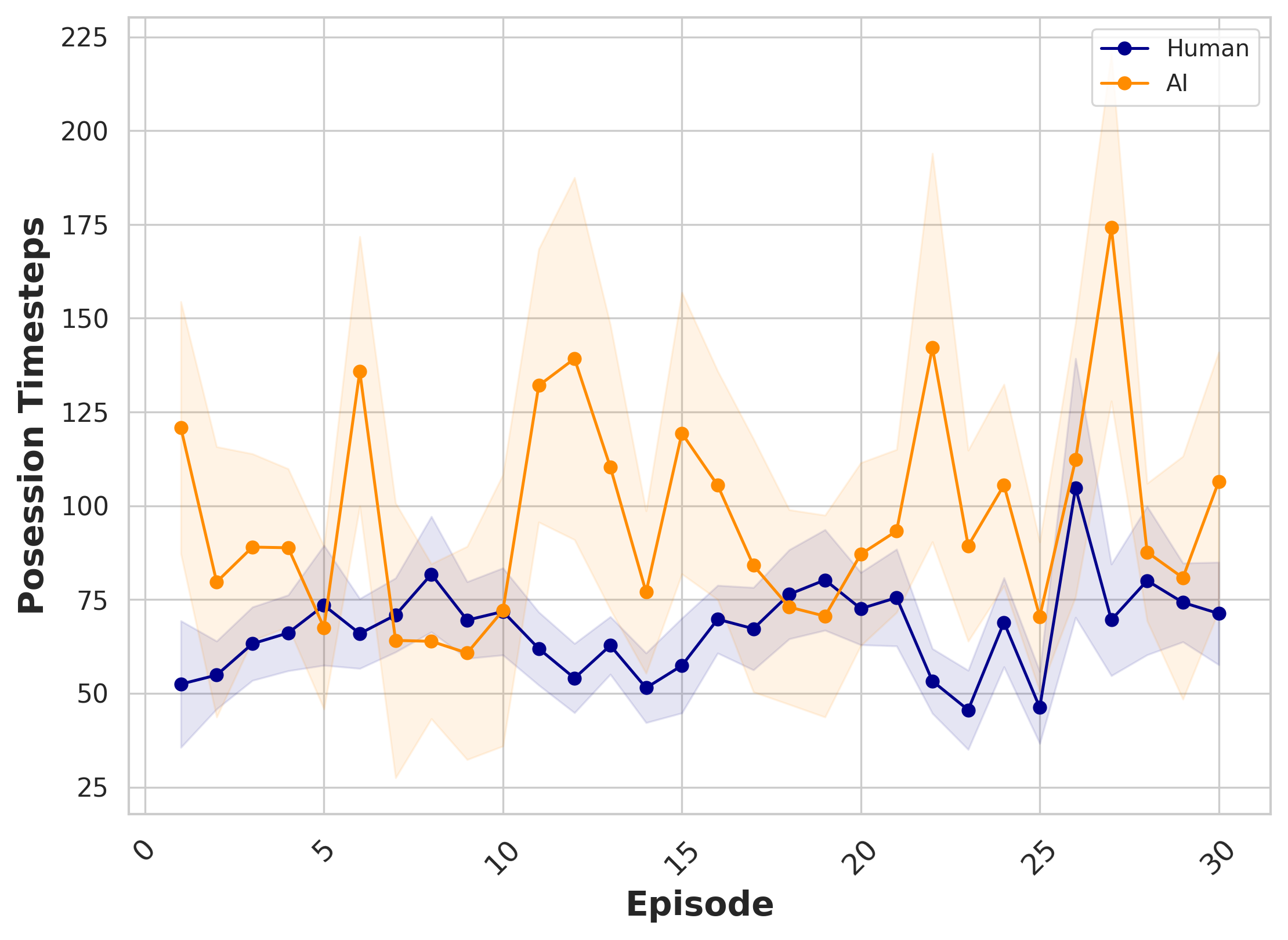}
    \caption{The average number of timesteps per episode in the Human-AI study where the ball is in possession of the human or AI. Possession is defined by the ball being on the corresponding side of the middle post. The AI agents have higher possession compared to the human counterparts. \red{Shaded regions represent standard error.}}
    \label{fig:slime_vb_possession}
\end{figure}

\paragraph{Results.} Results are shown in Figure~\ref{fig:slime_vb_ep_len}. Because Slime Volleyball is a zero-sum game, cumulative score is not an informative measure; instead, we report the episode length as a proxy for player skill. Longer episodes indicate a greater ability to keep the ball in play.  In these short sessions, we observed little-to-no improvement over time. Human–AI pairs maintained longer rallies than Human–Human pairs, reflecting the trained agent’s higher baseline skill. Indeed, AI–AI matches often reached the maximum length of 3,000 steps. Figure~\ref{fig:slime_vb_possession} further shows that the AI maintained higher “possession,” keeping the ball on its side for longer periods. \red{The relatively flat episode lengths reflect the difficulty of the task: Slime Volleyball is a fast-paced motor task requiring precise timing with unintuitive physics, and 30 episodes provides limited practice for substantial skill improvement. Human-AI pairs sustain longer rallies because the trained agent reliably returns the ball, keeping it in the air for longer. Human-human pairs' episode length does not benefit from the high skill of the AI.} Although average performance remained \red{relatively low}, some participants substantially exceeded baseline, with the longest Human–AI match lasting 1,169 steps and the longest Human–Human match 538 steps.

The Slime Volleyball study demonstrates that \mug is not limited to grid-based environments created with \cogrid. The support for fast-paced, physics-based tasks illustrates \mug's flexibility for running studies with rich interaction dynamics and complexity, broadening the scope of potential investigation. 
\section{Discussion}

Our contributions aim to expand the ecosystem of platforms for multi-agent research. Existing environments and libraries \citep[e.g.,][]{terry2021pettingzoo, agapiou2022melting, carroll_utility_2020} have provided standardized testbeds for AI agents, but they aren't designed to prioritize customization or accessibility. Similarly, work in human-AI teaming has been carried out using non-standard or bespoke experimental platforms that have significant barriers to entry, preventing researchers from extending them or investigating research questions with modern AI. \cogrid and \mug help to solve these problems by enabling simple customization of multi-agent environments and uniquely providing a standardized method to translate simulation environments into web-based experiments, respectively. Our framework complements existing platforms by allowing researchers to move beyond isolated simulations and move to empirical evaluations with human participants. 

Our work on \cogrid extends the popular Minigrid library to the multi-agent domain, allowing researchers to easily define their own environments with minimal overhead and barriers to entry. It modularizes the components of Minigrid to enable arbitrary manipulation of environments and customizable execution logic in grid-based environments. We illustrated \cogrid's functionality through a case study of implementing Overcooked \citep{carroll_utility_2020} in \cogrid, demonstrating the streamlined process by which new objects and interactions can be added to an environment. Multi-agent environments, like those that can be designed with \cogrid, provide opportunities to investigate interactions between agents---which is particularly important as AI is introduced into our social world. 

Where other libraries are primarily focused on benchmarks and standardized environments (e.g., PettingZoo \citep{terry2021pettingzoo}, Melting Pot \citep{agapiou2022melting}, Overcooked-AI \citep{carroll_utility_2020}), \cogrid emphasizes and prioritizes modularity, accessibility, and customizability. This focus situates \cogrid as a library that will support researchers in asking questions about novel interactions not represented by current benchmarks. The primary \red{limitation} of \cogrid is the grid-based nature of the environments. Although substantial complexity can arise in grid-based interactions, it excludes physics-based or continuous control settings that may be of interest in other domains. 

In addition to \cogrid, \mug is a unique contribution that establishes a streamlined process to translate simulation environments to interactive experiments in the browser. Where environments would typically have to be reimplemented to run in an interactive experiment, \mug provides a tool for researchers to produce a \red{multiplayer} game directly from a Gymnasium or PettingZoo environment. Through demonstrations with Overcooked and Slime Volleyball, we showed how \mug facilitates both Human-AI and Human-Human experiments without requiring bespoke reimplementations. These libraries lower technical barriers and allow researchers to study human-AI interaction in an efficient and scalable manner.   

\red{A significant limitation is that \mug does not currently support explicit communication between participants or between humans and AI. All interaction is implicit, mediated through actions and observations in the shared environment. There is no built-in text chat or structured messaging system, which limits applicability to studies that require communication. Future work will address this by incorporating communication capabilities into \mug in the form of text-based chat. Furthermore, \mug is currently aimed at interactions within simulation environments, but could be extended to support complex interactions that include communication, demonstration, preference elicitation, and other advanced features.}

Despite these limitations, there are significant opportunities enabled by this framework. With \cogrid, researchers can prototype novel multi-agent scenarios or dilemmas to answer their particular questions rather than relying on pre-defined settings. With \mug, such environments can be brought online to study how humans interact with the AI that was trained in them to answer questions on perception, preferences, and behavior in human-AI interaction.  

Through the introduction of \cogrid and \mug, we have demonstrated a simple setting to develop multi-agent simulation environments and a framework that streamlines the process of developing interactive experiments in simulation environments. The goal of this work is to support research in multi-agent systems, particularly in human-AI interaction and complementarity. By providing tools that support the rapid development and testing of ideas, we hope to contribute to progress in building autonomous agents that better serve the humans with whom they interact in the real world.

\section*{Open Practices Statement}

All source code, documentation, and tutorials for \cogrid and \mug are available at \url{cogrid.readthedocs.io} and \red{\url{multi-user-gymnasium.readthedocs.io}}, respectively. The scripts to recreate the experiments are available in the respective code repositories.

\section*{Declarations}

\paragraph{Funding}

This research was supported by the Defense Advanced Research Projects Agency and was accomplished under Grant Number W911NF-20-1-0006 and by the NSF AI Institute for Societal Decision Making (AI-SDM), Award No. 2229881.

\paragraph{Conflicts of Interest}

The authors have no competing interests to declare that are relevant to the content of this article.

\paragraph{Ethics Approval}

Ethics approval was received by the Institutional Review Board at Carnegie Mellon University under the project \emph{Building Human-Machine Shared Mental Models}, study \emph{STUDY2021\_00000324}.

\paragraph{Consent to Participate}

Informed consent was obtained from all individual participants.

\paragraph{Consent for Publication}

The participants gave their consent for the publication of their behavioral data.  

\paragraph{Availability of Data and Materials}

Data are available through \url{cogrid.readthedocs.io} and \red{\url{multi-user-gymnasium.readthedocs.io}}.

\paragraph{Code Availability}

All source code and documentation are available through \url{cogrid.readthedocs.io} and \red{\url{multi-user-gymnasium.readthedocs.io}}.

\printbibliography

% \printbibliography
% \bibliographystyle{apacite}
% \bibliography{ref2}

\ifdefined\includeappendix
\appendix
\section{Overcooked Environment \& Task} \label{sec:overcooked_appendix}

The full implementation of our adaptation of the Overcooked-AI environment \citep{carroll_utility_2020} is documented at \url{cogrid.readthedocs.io}. In this section, we provide an overview of the design of the environment and interactive task. All code to use the \cogrid Overcooked environment with \mug is available as an example in the \mug repository, linked at: \url{multi-user-gymnasium.readthedocs.io}.

\subsection{\cogrid Overcooked}

\red{The Overcooked environment is implemented by subclassing \pyinline{CoGridEnv} and defining the task-specific components through \cogrid's registration system. Each component---objects, rewards, and features---is a Python class that is registered at initialization. The environment layout is specified via an ASCII grid encoding in the configuration file. Full source code and implementation details are available in the online documentation at \url{cogrid.readthedocs.io}.}

\red{Listing~\ref{listing:onion_snippet} shows the \pyinline{Onion} object, illustrating how environment objects are defined. The \pyinline{@register_object_type} decorator registers the class within the environment scope. Interaction capabilities are declared as class attributes using the \pyinline{when()} descriptor---here, \pyinline{can_pickup = when()} indicates that onions can always be picked up. Each object specifies a unique \pyinline{char} attribute used for ASCII layout encoding.}

\begin{listing}[!ht]
    \begin{mintedbox}{python}
@register_object_type("onion", scope="overcooked")
class Onion(GridObj):
    color = Colors.Yellow
    char = "o"
    can_pickup = when()
    \end{mintedbox}
    \caption{\red{The \pyinline{Onion} object. The \pyinline{@register_object_type} decorator registers the class and the \pyinline{when()} descriptor declares that onions can be picked up. Additional objects (\pyinline{Plate}, \pyinline{OnionSoup}, \pyinline{DeliveryZone}, etc.) follow the same pattern.}}
    \label{listing:onion_snippet}
\end{listing}

\red{Rewards are defined using the \pyinline{InteractionReward} base class. Instead of writing complex reward logic, subclasses declare the triggering conditions as class attributes: the \pyinline{action} the agent must perform, the item it must \pyinline{hold}, and the object it must \pyinline{face}. The base class handles all condition checking and reward broadcasting. Listing~\ref{listing:reward_snippet} shows the delivery reward, which triggers when an agent performs a pickup/drop action while holding an \pyinline{OnionSoup} and facing a \pyinline{DeliveryZone}. For more complex conditions (e.g., checking pot capacity or cooking status) subclasses can override an \pyinline{extra_condition()} method to specify additional conditions.}

\begin{listing}[!ht]
    \begin{mintedbox}{python}
class OnionSoupDeliveryReward(InteractionReward):
    action = "pickup_drop"
    holds = "onion_soup"
    faces = "delivery_zone"
    \end{mintedbox}
    \caption{\red{The soup delivery reward using the declarative \pyinline{InteractionReward} base. Only class attributes are needed: the action, what the agent holds, and what it faces.}}
    \label{listing:reward_snippet}
\end{listing}

\begin{listing}[!ht]
    \begin{mintedbox}{python}
class OnionInPotReward(InteractionReward):
    action = "pickup_drop"
    holds = "onion"
    faces = "pot"

    def extra_condition(self, mask, prev_state,
                        fwd_r, fwd_c, reward_config):
        # Check pot has capacity for another onion
        return condition_satisfied
    \end{mintedbox}
    \caption{\red{A reward with an additional condition. The \pyinline{extra_condition()} override narrows the trigger mask to check pot capacity and ingredient compatibility.}}
    \label{listing:reward_extra_condition}
\end{listing}

\red{Features for the observations follow a similar pattern, where users can define custom features by subclassing the \pyinline{Feature} base class, and generate a function that maps from the current state of the environment to specific view of the environment (e.g., a one-hot encoding of which item an agent is holding). The full details and implementations are shown in the online documentation.}

\red{For the \pyinline{Overcooked} environment, the observation features roughly correspond to those used by \cite{carroll_utility_2020}.} For each agent $j$, we calculate the following features. The observation is then the concatenation of all agents' feature arrays:
\begin{itemize}
    \item Agent $j$'s direction as a one-hot encoding.
    \item Agent $j$'s inventory as a one-hot encoding of the possible inventory objects.
    \item A multi-hot indicator of whether agent $j$ is adjacent to a counter.
    \item Agent $j$'s distance to the closest of each \pyinline{Onion}, \pyinline{Plate}, \pyinline{PlateStack}, \pyinline{OnionStack}, \pyinline{OnionSoup}, and \pyinline{DeliveryZone}.
    \item Agent $j$'s pot features, consisting of:
        \begin{enumerate}
            \item an indicator of whether the pot is reachable,
            \item a one-hot representation of the pot status, which can be empty, cooking, or ready;
            \item the number of onions in the pot,
            \item the number of cooking timesteps remaining for the pot,
            \item an array of the row and column distances to the pot,
            \item the row and column location of the pot.
        \end{enumerate}
    \item Agent $j$'s distance to the other chef.
    \item Agent $j$'s row and column position in the grid.
\end{itemize}

\subsection{Training a Reinforcement Learning Agent} \label{sec:training_rl_cogrid}

We train a reinforcement learning agent in the \pyinline{Overcooked} environment using RLlib \citep{liang2018rllib} and the PPO algorithm. \red{A dish delivery reward of 1.0 is given when a dish is delivered, and agents also receive 0.1 reward when an onion is placed in a pot and 0.3 when a dish is plated.} \red{The full training script, including all hyperparameters, is available in the online documentation and source code. We also provide a complete example for training with full hardware acceleration (using the JAX backend). The trained policy delivers an average of roughly 7.5 dishes per episode when controlling both agents.}

% The training progression is shown in Figure~\ref{fig:oc-rl-training}. We train the reinforcement learning policy for roughly 250 million environment timesteps using self-play, which means that the single policy controls the behavior of both agents in the Overcooked environment. 

% \begin{figure}
%     \centering
%     \includegraphics[width=\textwidth]{figures/overcooked_training.png}
%     \caption{The training progression of the Overcooked reinforcement learning agent. Shaded regions represent the standard error in performance over 20 episodes.}
%     \label{fig:oc-rl-training}
% \end{figure}

\subsection{Overcooked in \mug} \label{sec:overcooked_in_ig}

The full source code is available in the examples of the \mug repository, documented at: \url{multi-user-gymnasium.readthedocs.io}. The online documentation provides all implementation details and configurations to launch Overcooked in \mug.

\section{Slime Volleyball Environment \& Task} \label{sec:slime_vb_training_appendix}

The Slime Volleyball environment we use was originally developed by \cite{slimevolleygym}. We adapted their implementation to conform to the Gymnasium API, but did not make any major alterations to the environment dynamics. The code for the Gymnasium formatted environment is available at \url{https://github.com/chasemcd/slimevolleygym}. 

The observation $o_t\in\mathcal{O}$ is a size 12 vector with the $(x,y)$-position and $(x,y)$-velocity for each player and the ball.

We use a modified training regime for Slime Volleyball with a self-play curriculum. The agent trains against a fixed version of itself, which is updated every time the average reward against that agent exceeds 0.5---agents receive a reward of -1 for losing, 0 for ties, and 1 for winning. \red{The full training script, including all hyperparameters, for Slime Volleyball are provided in the \mug documentation and source code. Over the course of training, the policy is able to achieve an average episode length of nearly the maximum of 3,000 steps in the Slime Volleyball environment.}

% The training progression, in terms of episode length in self-play, is shown in Figure~\ref{fig:sv_training}. Over the course of 350 million environment steps, the average episode length reaches nearly the maximum of 3,000 in the Slime Volleyball environment. 

% \begin{figure}
%     \centering
%     \includegraphics[width=0.75\textwidth]{figures/slime_vb_training_progression.png}
%     \caption{Episode length over time during training for the Slime Volleyball reinforcement learning agent. The policy is trained for roughly 350 million environment steps. Significant variation occurs when the agent that the policy is training against is replaced, which occurs every time an average reward of 0.5 is earned against it.}
%     \label{fig:sv_training}
% \end{figure}

\subsection{Slime Volleyball in \mug}

 The full configuration and implementation details to launch Slime Volleyball in \mug are documented at \red{\url{multi-user-gymnasium.readthedocs.io}}. 
\fi

\end{document}